\documentclass[review]{elsarticle}
\usepackage{amsmath}
\usepackage{amssymb}
\usepackage{hyperref}
\usepackage{url}
\usepackage[]{subfig}
\usepackage{textgreek}
\newcommand{\SI}[2]{\ensuremath{{#1}\;{\mbox{#2}}}}
\newcommand{\Nuc}[2]{\ensuremath{^{#2}\mbox{#1}}}
\usepackage{booktabs}

\begin{document}
\begin{frontmatter}
\title{In-situ characterization of the Hamamatsu R5912-HQE photomultiplier tubes used in the DEAP-3600 experiment}

\author[j]{P.-A.~Amaudruz}
\author[c]{M.~Batygov}
\author[a]{B.~Beltran}
\author[a]{C.\,E.~Bina}
\author[j]{D.~Bishop}
\author[e]{J.~Bonatt}
\author[h]{G.~Boorman}
\author[e,b]{M.\,G.~Boulay}
\author[e]{B.~Broerman}
\author[g]{T.~Bromwich}
% ...

\address[a]{Department of Physics, University of Alberta, \\Edmonton, Alberta T6G 2R3, Canada}
\address[b]{Department of Physics, Carleton University, \\Ottawa, Ontario, K1S 5B6, Canada}
\address[c]{Department of Physics and Astronomy, Laurentian University, \\Sudbury, Ontario, P3E 2C6, Canada}
\address[d]{Instituto de F\'isica, Universidad Nacional Aut\'onoma de M\'exico,\\ Apartado Postal 20-364, M\'exico D. F. 01000, Mexico}
\address[e]{Department of Physics, Engineering Physics, and Astronomy, Queen's University, \\Kingston, Ontario K7L 3N6, Canada}
\address[f]{SNOLAB, Lively, Ontario P3Y 1M3, Canada}
\address[g]{Department of Physics and Astronomy, University of Sussex, \\Sussex House, Brighton, East Sussex BN1 9RH, United Kingdom}
\address[h]{Department of Physics, Royal Holloway, University of London, \\Egham Hill, Egham, Surrey TW20 0EX, United Kingdom}
\address[j]{TRIUMF, Vancouver, British Columbia V6T 2A3, Canada}
\address[k]{Department of Physics, Technische Universit\"at M\"unchen, \\80333 Munich, Germany}
\address[l]{Canadian Nuclear Laboratories Ltd (former Atomic Energy of Canada Ltd), Chalk River Laboratories, \\Chalk River, K0J 1P0 Canada}
\fntext[dec]{Deceased.}
\cortext[cor1]{Corresponding author. Email: deap-papers@snolab.ca}

\begin{abstract}
The Hamamatsu R5912-HQE photomultiplier-tube (PMT) is a novel high-quantum efficiency PMT. It is currently used in the DEAP-3600 dark matter detector and is of significant interest for future dark matter and neutrino experiments where high signal yields are needed.
 
We report on the methods developed for in-situ characterization and monitoring of DEAP's 255 R5912-HQE PMTs. This includes a detailed discussion of typical measured single-photoelectron charge distributions, correlated noise (afterpulsing), dark noise, double, and late pulsing characteristics.
The characterization is performed during the detector commissioning phase using laser light injected through a light diffusing sphere and during normal detector operation using LED light injected through optical fibres.
\end{abstract}


\end{frontmatter}

\section{Introduction} \label{sect:intro}

The DEAP-3600 detector was built to detect interactions between WIMP (weakly interacting massive particles) dark matter and argon nuclei with a projected sensitivity of $10^{-46}$cm$^2$ to the spin-independent dark matter-nucleon interaction cross section \cite{Kuzniak:2017dt}. DEAP-3600 is a single-phase liquid argon (LAr) detector with a fiducial mass of \SI{1}{tonne}. The sole signal channel is scintillation light from particle interactions in the LAr, which is registered by 255 Hamamatsu R5912 high quantum efficiency photomultiplier tubes (PMTs).

In order to reach the projected sensitivity, the experiment relies on pulse shape discrimination (PSD) and event position reconstruction. PSD is used to separate the nuclear-recoil signal events from the large rate of electromagnetic background events due to \Nuc{Ar}{39} beta decays. Position reconstruction is used to define an ultra-clean fiducial region of the detector by rejecting backgrounds due to alpha particles entering the LAr from the inner detector surface.

The power of PSD and position reconstruction to reject backgrounds is tied to accurate counting of photoelectrons in each particle-interaction event. Accurate counting requires understanding how PMT instrumental effects such as afterpulsing, late pulsing, and dark noise modify the observed LAr scintillation pulse shape. Ongoing calibration, characterization, and monitoring of each of the 255 PMTs is thus necessary to reach the project's target sensitivity. 


We report on the calibrations possible with the detector hardware and on the methods that have been developed to characterize and to monitor PMT properties in-situ. Results are discussed for a representative PMT. This is the first time the afterpulsing characteristics and single photo-electron charge distribution shape for these PMTs is reported on. Ensemble results across the 255 PMT array can be found in \cite{detectorpaper}. A discussion of the implications of the PMT properties measured using these methods will be part of upcoming papers.



\section{The DEAP-3600 detector} \label{sect:setup}
\subsection{Detector description} \label{sec:d3detector}
A detailed description of the DEAP-3600 detector can be found in \cite{detectorpaper}. We give only a brief overview here.
\begin{figure}[htbp] 
\centering
	\subfloat[]{\includegraphics[width=0.4\textwidth]{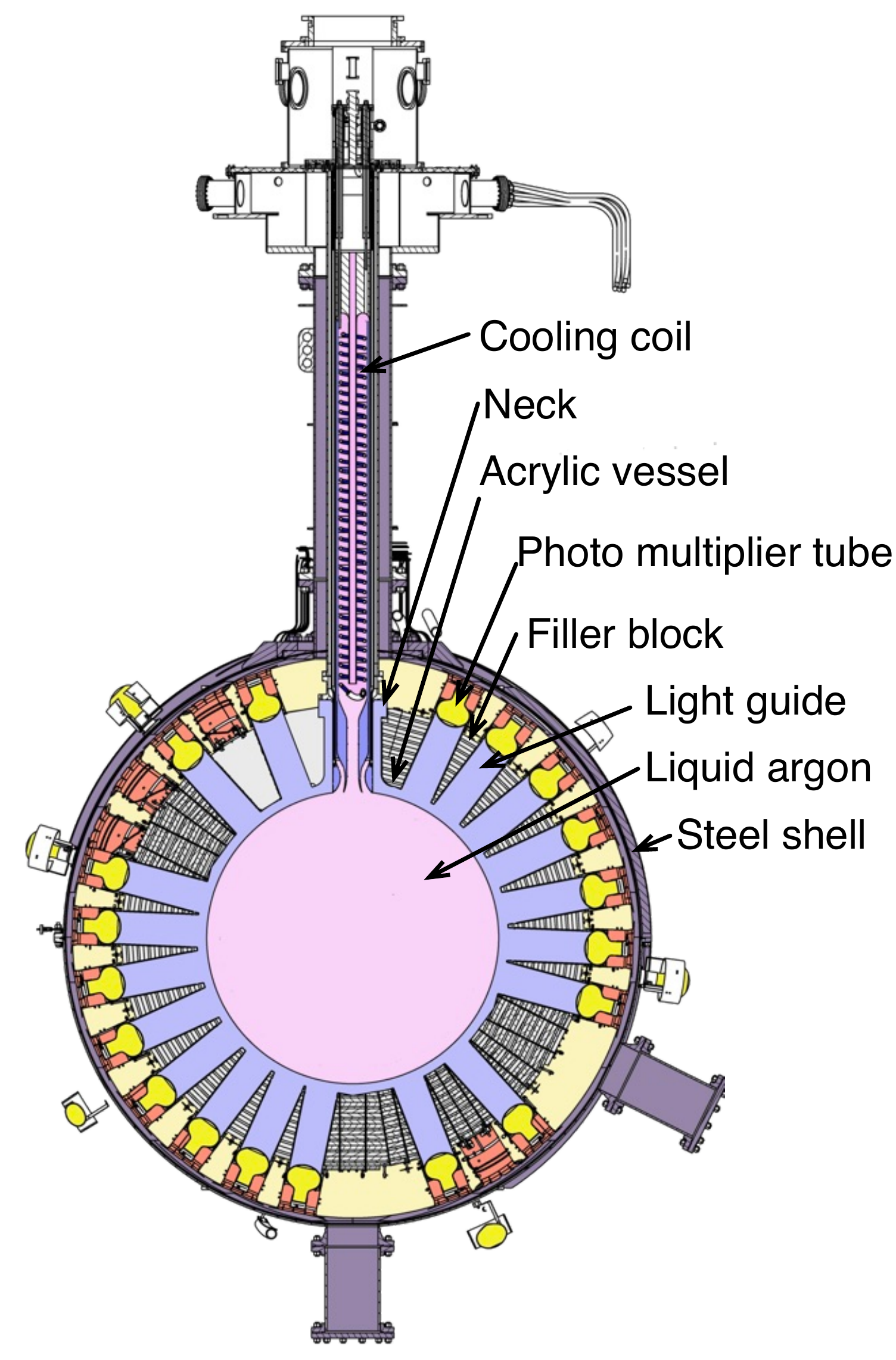}}
     \subfloat[]{\includegraphics[width=0.55\textwidth]{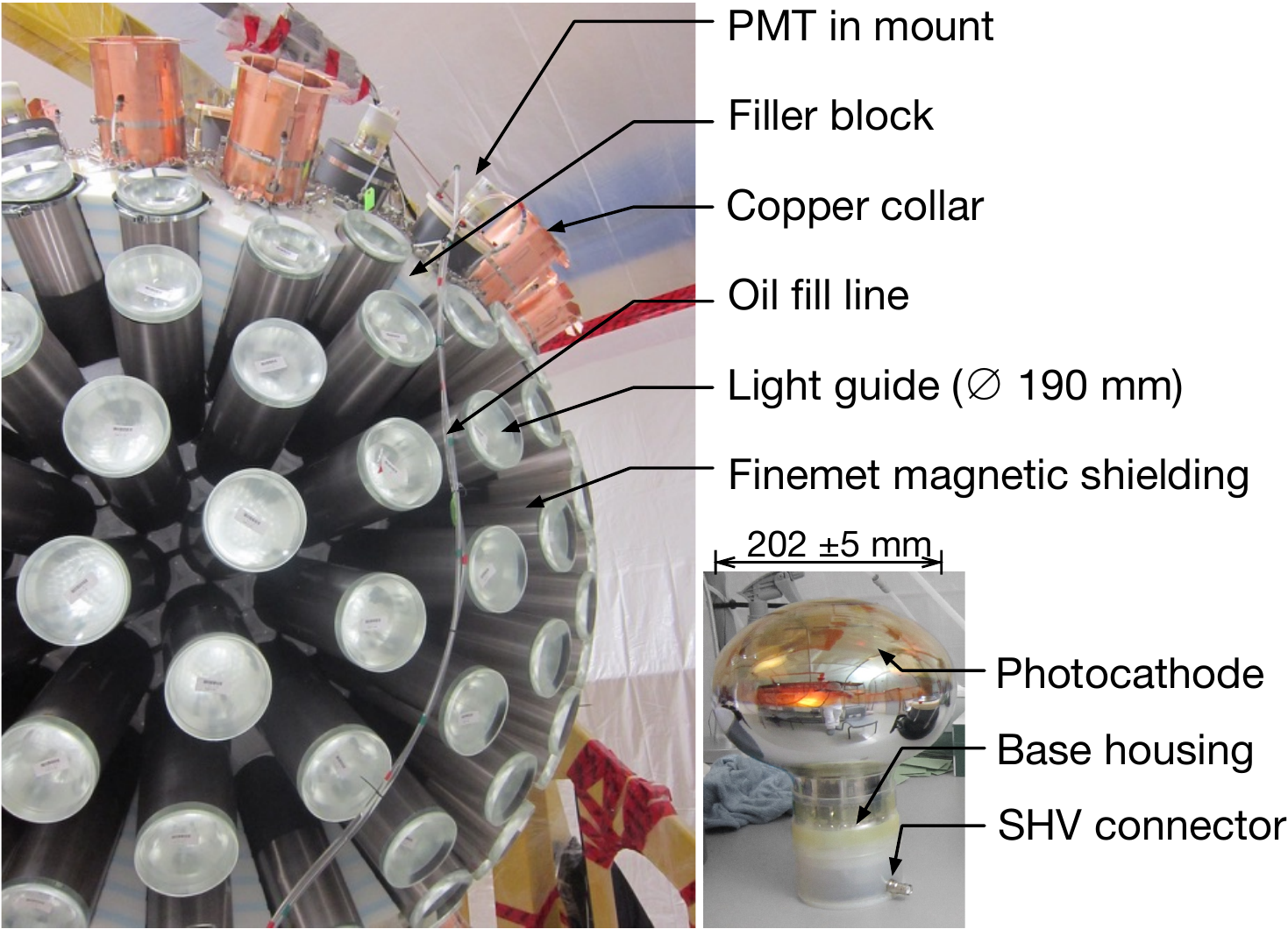}}
     \caption[]{a) Design drawing of the DEAP-3600 detector. b) Photo of the DEAP-3600 detector during assembly. The left photo shows the light guides protruding from the acrylic vessel. The light guides and vessel wrapped in reflector materials are then covered in black photon-absorbing material, and the ends of the light guides are further surrounded by Finemet magnetic shielding foil. Several PMTs are installed on their light guides and one PMT assembly is being filled with optical coupling oil. Some PMTs are already enclosed in their copper and Finemet collars. The right picture shows a bare Hamamatsu R5912 8" PMT (the dimension is shown in the figure with the tolerance across all PMTs). The PMT base is sealed in a plastic housing and the housing is filled with two component soft silicone gel. The combined bias-voltage supply and signal read-out connector protrudes from the housing.}
     \label{fig:avassembly}
\end{figure}

At the centre of the DEAP-3600 detector is a spherical volume \SI{170}{cm} in diameter filled with \SI{3.6}{tonnes} of LAr. Liquid argon emits scintillation light with a wavelength of \SI{128}{nm} in response to particle interactions. The scintillation light travels through the argon volume until it reaches the surface of the acrylic containment vessel. The inside acrylic surface is coated with the  organic wavelength shifter TPB\cite{TPBpaper}, which shifts the scintillation light to the blue spectral region. The wavelength-shifted light is transmitted to the PMTs through a total of \SI{50}{cm} of acrylic in the form of an acrylic vessel (AV) and acrylic light guides (LGs). The 255 cylindrical light guides protrude radially from the acrylic vessel and at \SI{19}{cm} in diameter they provide 76\% surface coverage. The space between the light guides is filled by blocks (``filler blocks'') made from alternating layers of polyethylene and Styrofoam SM. 

\begin{figure}[htbp] 
     \includegraphics[width=\textwidth]{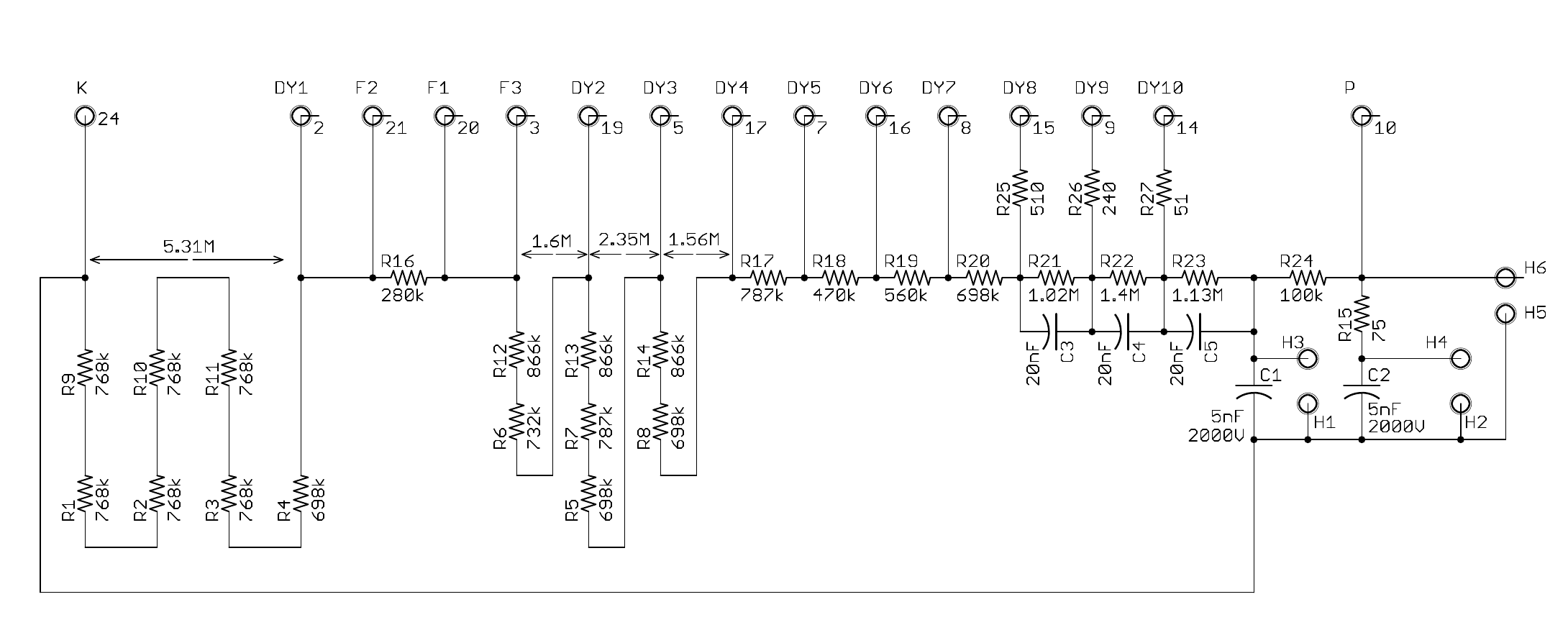}
     \caption[]{Circuit diagram for the PMT bases. The dynode voltages are divided in a tapered manner, and the anode has a \SI{75}{$\Omega$} terminating resistor to minimize reflections. The labels in the figure denote: K - cathode, DYn - dynodes, Fn - focusing electrodes, P - anode, and Hn - solder pads on circuit board.}
     \label{fig:bases}
\end{figure}

A Hamamatsu R5912 high quantum efficiency PMT is optically coupled, using silicone oil, to the end of each light guide. The diameter of the light guides is approximately equal to the diameter of the PMT bulbs. Fig.~\ref{fig:avassembly} shows the detector during installation of the PMTs. Each PMT is surrounded by a Finemet foil collar within a copper collar. The Finemet shields residual magnetic fields not compensated by the compensation coils (described later) while the copper prevents large temperature gradients across the length of the PMT.  PMTs are labelled with an ID number from 0 through 254.

Temperature sensors are installed on the copper collars of 16 PMTs distributed across the detector.  The measurements from these sensors are used to quantify the PMT temperature for the studies that follow. The temperature of PMTs without a sensor is estimated from the reading of the sensor closest to them.  

The inner detector, consisting of the argon target inside the acrylic vessel, light guides, insulation material, and PMT assemblies, is sealed inside a stainless steel sphere. This sphere is suspended in an \SI{8}{m} diameter water shielding tank (to veto passing muons and to moderate neutrons) and surrounded by four magnetic compensation coils, which complete the magnetic shielding of the PMTs.

The experiment is located \SI{2}{km} underground in the Cube Hall at SNOLAB\cite{snolab} in Sudbury, Canada.

\subsection{PMT properties and operating conditions} \label{sec:phases}

The R5912-HQE PMT has an \SI{8}{inch} borosilicate glass bulb with a bialkali photocathode. The dynodes are arranged in the \emph{box type} structure with 10 amplification stages and a typical gain of $1\cdot10^7$. The quantum efficiency as stated by the producer is approximately 23\% at \SI{400}{nm}, and the active photocathode area is approximately \SI{530}{cm$^2$}\cite{Hamamatsu:r5912}.

The PMTs used in DEAP-3600 have a tapered base-circuit as shown in Fig.~\ref{fig:bases} and the bases are individually sealed in a plastic housing filled with two-component soft silicone gel. The PMTs are operated at positive high voltage (HV) mainly to allow using a single cable for signal read-out and HV supply on each PMT. This helps to reduce the radioactivity level near the detector and alleviates space constraints when routing the cables. The bias voltages are set independently for each PMT such that the PMT array has a uniform gain. Typical bias voltages are between \SI{1400}{V} and \SI{1800}{V}.

The PMTs were operated under an atmosphere of gaseous N$_2$ with an admixture of 10\%~CO$_2$. Before colling, the environment was changed to pure N$_2$. Ambient magnetic fields are actively and passively shielded (see Sect.~\ref{sec:d3detector}). Initial optical calibrations were performed while the PMTs were at an ambient temperature of \SI{20}{$^\circ$C}. At that time, the AV was at vacuum and the water shielding tank was empty. The PMTs cooled down to \SI{10}{$^\circ$C} when the shield tank was filled with chilled water, and reached final temperatures of \SI{-35}{$^\circ$C} to \SI{+5}{$^\circ$C} once the AV was filled with LAr.

All data were taken well after the PMTs' dark noise rate had settled down at full bias voltage.

\subsection{The data acquisition system}\label{sec:DAQ}

Each PMT is connected to the data acquisition system (DAQ) through a single \SI{20}{m} RG-59 cable that carries both the kV-level bias voltage and the mV-level signal. The DAQ electronics are located outside the water tank. Signal conditioning boards (SCBs) decouple the signal from the bias voltage, and stretch and amplify the pulses in preparation for digitization. Each SCB channel is read out by a \SI{250}{MS/s}, 12 bit CAEN V1720 digitizer channel. 

In typical operating conditions, the digitizers save only the parts of the waveform that contain a pulse through zero length encoding (ZLE). Pulse detection is based on the waveform crossing a fixed threshold at \SI{2.44}{mV}, which corresponds to approximately 10\% of the single-photoelectron (SPE) pulse height. The DAQ can also record the full waveform (FWF); this mode is used sparingly as the data volume quickly becomes prohibitive. 

In optical calibration mode, the DAQ system triggers both light emission and data recording using a periodic \SI{0.5}{kHz} to \SI{2}{kHz} trigger. Each light flash represents one detector event, and due to the combined trigger, each event has one light flash at a fixed position in the recorded waveform. All PMTs are read out for each event.
Optical calibration data is taken with one of three acquisition settings summarized in Tab.~\ref{tab:daqsettings}

\begin{table}[htp]
\caption{DAQ settings for the data used in this paper. $\Delta T$ is the total length of the recorded waveform and $t_{light}$ is the time within the waveform where the signal from the triggered light source is expected. The last column denotes the data format as described in the text.}
\begin{center}
\begin{tabular}{|c|c|c|c|}
ID &  $\Delta T$ & $t_{light}$ & Format   \\
 \hline
1 &  1~$\mu$s & $0.5~\mu$s & FWF  \\
 2 & 16~$\mu$s & $6.5~\mu$s &ZLE  \\
3 &  200~$\mu$s &$20~\mu$s &ZLE 
\end{tabular}
\end{center}
\label{tab:daqsettings}
\end{table}%

The FWF or ZLE data samples for each channel are sent to front-end computers and converted to the DEAP data structure in the ROOT data format. The data files are sent offsite for offline analysis.

For more information about the DEAP-3600 DAQ system, we refer the reader to \cite{detectorpaper}.

\section{Optical calibration}\label{sect:optcal}

The optical calibration systems consist of a near-isotropic light source (a \emph{laserball}) deployed at three different heights along the vertical axis of the empty acrylic vessel during commissioning, and a permanent LED light injection system with fibres near 20 of the PMTs and on the detector neck.

The occupancy of the PMT is defined as the number of times that PMT registers a pulse at the light injection time divided by the total number of laser or LED flashes: 
\begin{equation}
\mathcal O_{\text{PMT j}} = \frac{\text{\# light flashes detected by PMT j}}{\text{\# light flashes emitted}} \label{eq:occ}
\end{equation}

The data recorded from each PMT is a mixture of 0-PE or pedestal, 1-PE (single-photoelectron or SPE), and multi-PE pulses. The probability of observing an $N$~PE pulse follows Poisson statistics
\begin{equation}
\text{Poisson}(N,\lambda) = \frac{\lambda^N e^{-\lambda}}{N!} \label{eq:poisson}
\end{equation}
and the mean number of PE observed by PMT $j$, $\lambda_j$, is related to the occupancy as\footnote{Here we are neglecting threshold effects which suppress the number of detected pulses, and furthermore suppress SPE signals much more than double or multiple photoelectron signals.}

\begin{equation}
\lambda_j = -\ln(1-\mathcal O_{\text{PMT j}} ) \label{eq:gammaocc}
\end{equation}

It follows from Eqs.~\eqref{eq:poisson} and \eqref{eq:gammaocc} that at 5\% occupancy, the distribution of pulses is 95\% pedestal, 4.87\% SPE, and 0.13\% multi-PE. This is a good compromise between the purity of the SPE pulses and the fraction of time that a pulse actually occurs, and the light intensity of the laser or LED system was tuned to create this occupancy for most of the PMT calibrations. 

\subsection{LED light injection}
\begin{figure}[htbp] 
     \subfloat[]{\includegraphics[width=0.35\textwidth]{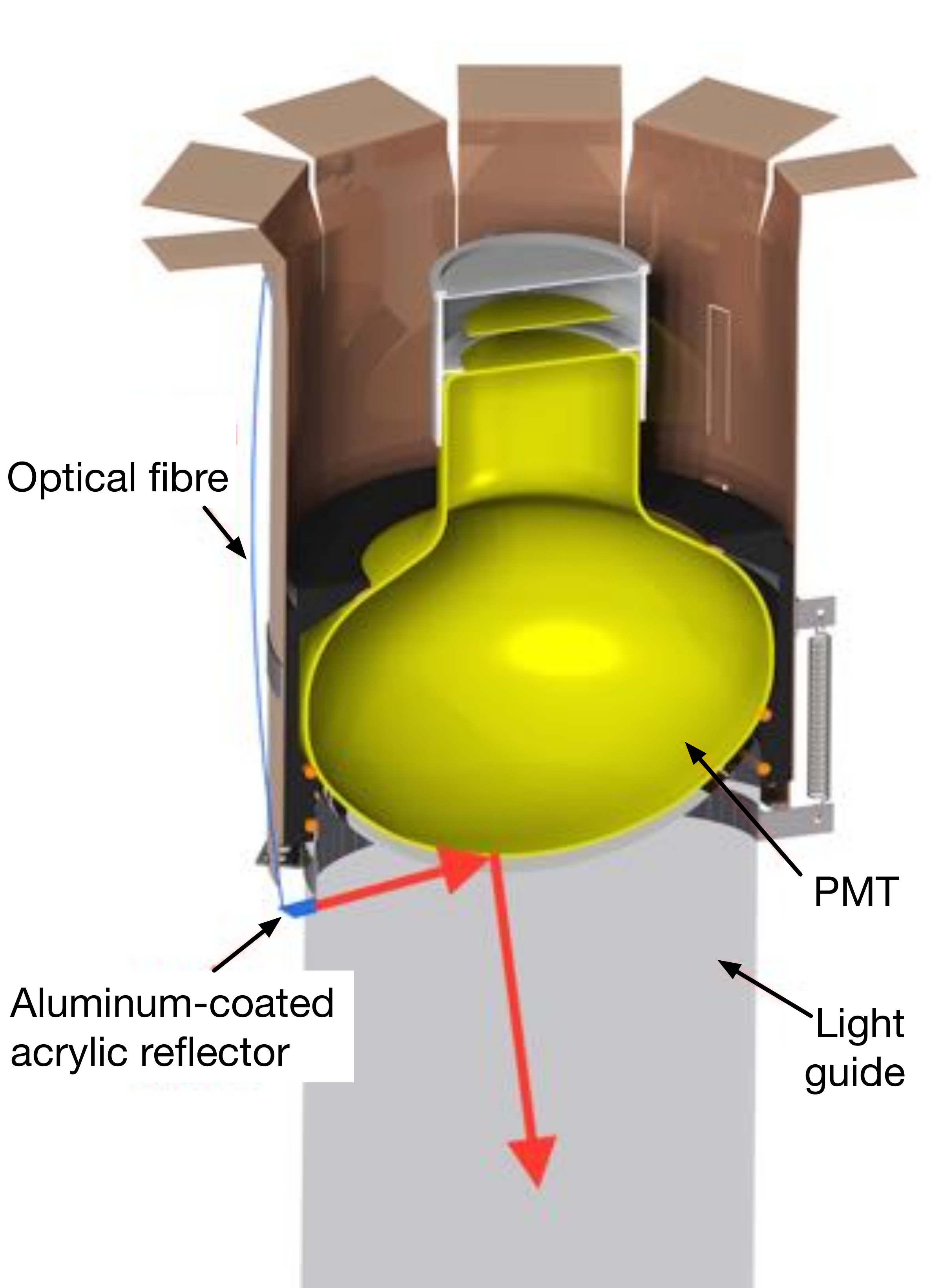}}
     \subfloat[]{\includegraphics[width=0.65\textwidth]{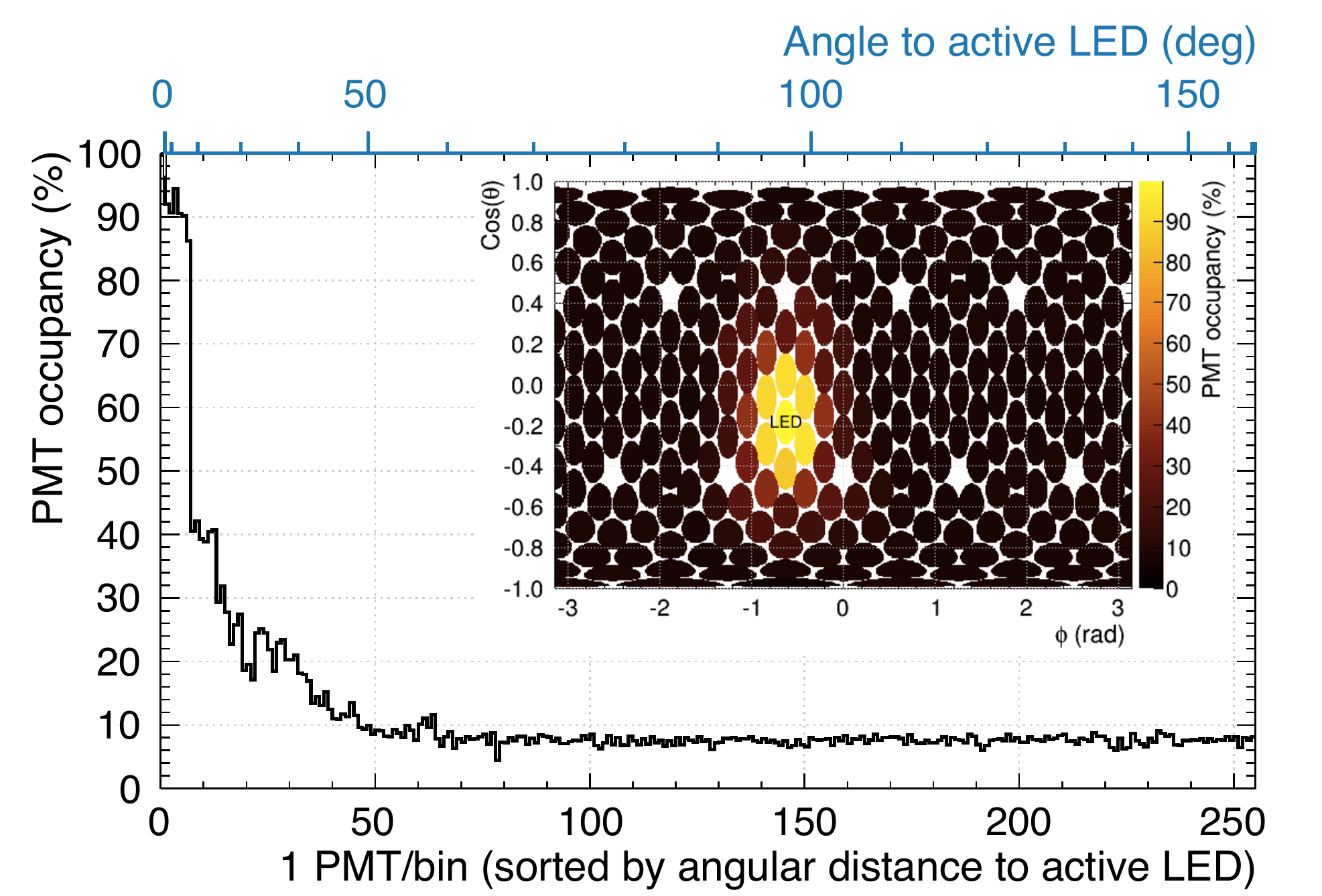}}
     \caption[]{a) Design drawing showing a PMT coupled to a light guide, with the optical fibre and fibre reflector belonging to the LED caibration system. The bold arrows indicate the path of the injected LED light. b) The light distribution, measured as PMT occupancy, for an LED calibration run, as function of the position of each PMT relative to the active LED being fired. The data is shown as a 1D distribution, and a 2D map, where each colored area represents the location of a PMT in the $\theta$ and $cos(\phi)$ spherical coordinates.}
     \label{fig:aarf}
\end{figure}

Aluminum-coated acrylic reflectors are bonded to 20 light guides at uniform spacing around the detector, and optical fibres are permanently glued to the reflectors\footnote{Two fibres are also coupled to reflectors installed in the detector neck, with the reflectors directing light into the AV sphere. These are not used in this work.}. The geometry is illustrated in Fig.~\ref{fig:aarf}. The reflectors direct the light injected into the fibres onto the PMT at the end of the light guide. PMTs that can be directly illuminated are called `fibre PMTs'. Much of the injected light from an individual fibre is detected by the fibre PMT, but we find that between 20\% and 25\% of the light is reflected off the fibre PMT and into the AV, so that the remaining PMTs also receive some of the injected light.\footnote{The percentage of reflected light is calculated by comparing the integrated light intensity measured by all PMTs to the intensity measured by the PMT that is directly illuminated}.

The light injection system is located outside the water tank. This system employs a \SI{435}{nm} LED board, driven by a Kapustinsky\cite{Kapustinsky:1985jr} pulser and triggered by the data acquisition system. The intensity of the LED on each channel is individually tuneable from zero up to several hundred photons emitted into the detector. 

LED light is injected into only a single fibre in any given LED calibration run. PMTs located more than \SI{50}{degrees} away from the illuminated fibre PMT all record approximately the same light intensity, with an occupancy variation between PMTs of less than 10\%. These PMTs see only the diffuse photons that travelled through the light guide and passed into the volume enclosed by the acrylic vessel, where they may be scattered by the TPB before passing back into the acrylic to reach a PMT.  PMTs within \SI{50}{degrees} of the LED have a higher occupancy than the others. These nearby PMTs receive photons that travel down the light guide that holds the reflector and then undergo total reflection at the acrylic surface, so that the photons never leave the acrylic. 

In order to obtain data at the same occupancy for all PMTs, two calibration datasets are recorded. Each dataset has one active fibre, and the fibres are chosen to be at opposite sites of the detector. The datasets are combined offline to obtain a full dataset where all PMTs have approximately the same occupancy.

Data for approximately 4~million LED light flashes were recorded typically every other day at acquisition setting 2 from Tab.~\ref{tab:daqsettings} with nominal 5\% occupancy on most PMTs. 10\% and 25\% median occupancy data were taken once a week. These datasets are used to monitor SPE charges and dark noise rates.

Every time the PMTs' temperature dropped by \SI{10}{K}, and at least once per month regardless of PMT temperature, 1~million 5\% occupancy events were recorded in acquisition setting 1 of Tab.~\ref{tab:daqsettings} for SPE charge distribution monitoring, and 10~million events were recorded in acquisition setting 3 of Tab.~\ref{tab:daqsettings} for afterpulsing studies.

\subsection{Laser light injection}
Laser light at \SI{450}{nm} wavelength was injected through the laserball into the TPB-coated AV while the AV was under vacuum. The laserball produces a sharp photon-arrival-time distribution due to excellent stability between flashes, so that the time of the initial pulse is known to better than \SI{1}{ns}.

5~million events were recorded with the laserball at the centre of the AV and 4 different light intensities in DAQ setting~2, and 1~million events in setting 1 (compare Tab.~\ref{tab:daqsettings}). The occupancy distribution across PMTs for the laserball is approximately flat so that a single dataset could be used to characterize all the PMTs at once. This data is used for the double and late pulsing studies.


\section{Measuring the SPE charge distribution}\label{sect:spe}
Due to statistical fluctuations in the amplification factor at each dynode, as well as physical effects such as backscattering, the charges measured in response to a single photoelectron released at the photocathode follow a wide distribution. The shape of this distribution determines the uncertainty in PE counting, and is used in detector simulations to accurately model pulse charges.

In this section, we study the SPE charge response using the low-light charge distribution.

\subsection{Measurement of the low-light charge distribution} 

The low-light charge distribution for each PMT is measured by creating a histogram of pulse charges from low-intensity light injected with the LED system. The charges shown here are always the charges above the mean level of the electronic baseline. As discussed in Sect.~\ref{sect:optcal}, this charge histogram is a superposition of the charge distributions from 0-PE, SPE, and multi-PE pulses, and the 0-PE charge distribution is centered at zero charge.
 
For FWF data, charges were sampled by integrating the digitized waveforms in a fixed window from \SI{24}{ns} before to \SI{44}{ns} after the LED flash time, without employing any pulse-finding algorithm. After amplification, an SPE pulse has a FWHM of about \SI{16}{ns}. A sample pulse is shown in the top panel of Fig.~\ref{fig:qfracBoth}. The \SI{68}{ns} integration time window was chosen to encompass shifts up to \SI{8}{ns} in the relative timing between channels, not yet calibrated out, and the time it takes photons to travel from the fibre end to the different PMTs (about \SI{9}{ns}). It also ensures that the larger and wider 2-PE pulses are not truncated. If the maximum of the pulse was within \SI{12}{ns} of the start or \SI{28}{ns} of the end of the integration window, it was discarded to avoid sampling partial pulse charges.

For ZLE data, a pulse finding algorithm was applied to the recorded sections of the waveform. The algorithm determined the peak time and total charge of the pulse, and pulses with a peak time from \SI{24}{ns} before to \SI{44}{ns} after the LED flash time were accepted in the charge histogram.

\begin{figure}[htbp] 
\centering
  \includegraphics[width=0.8\textwidth]{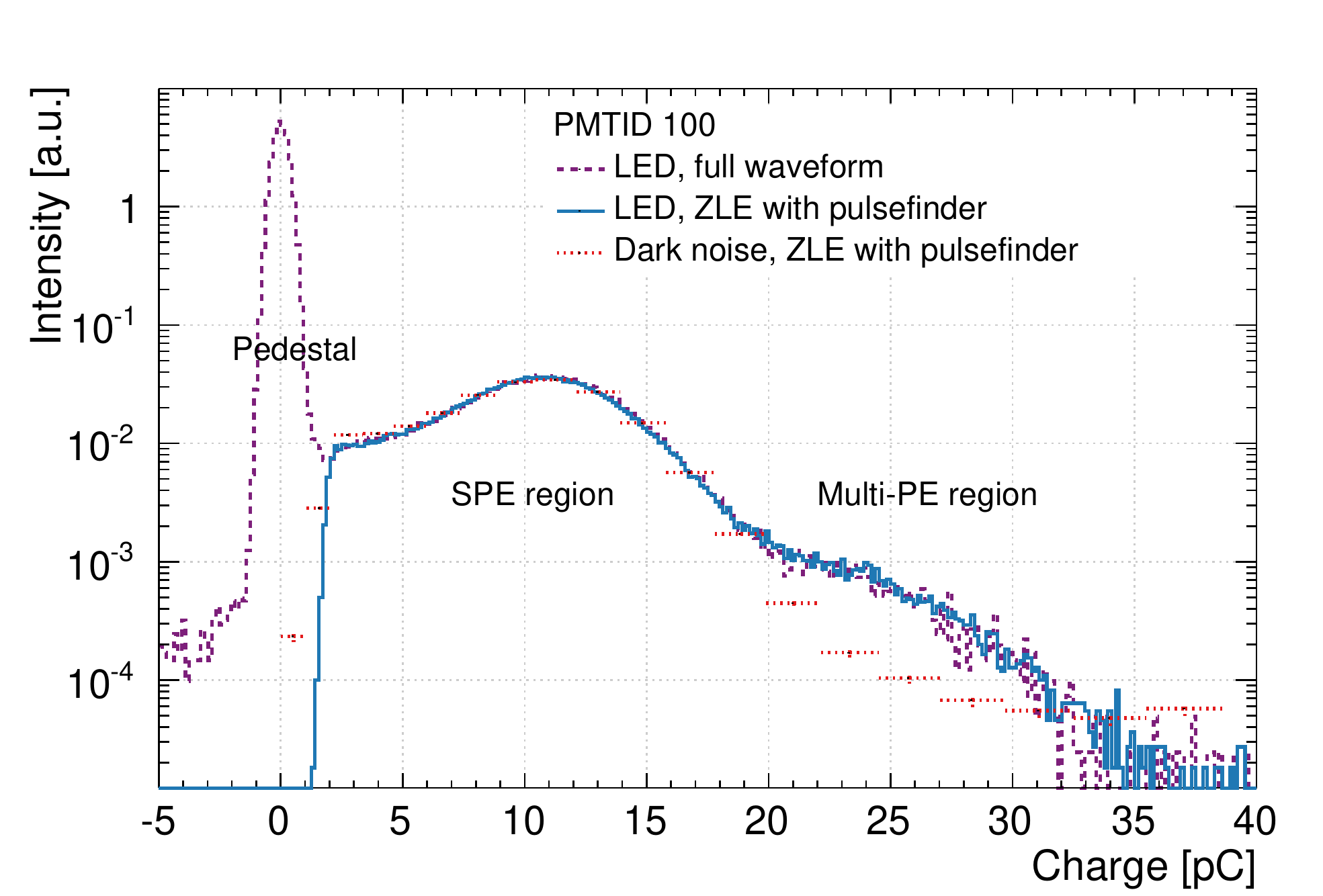}
  \caption{The low-light charge distribution for a sample PMT from
    an LED calibration run in FWF data taking mode where charges are
    sampled by fixed-window integration, and in ZLE mode with charges determined by the pulse
    finder. Also shown is the charge distribution from dark noise
    pulses. 
    The histograms are scaled to have the same SPE peak height.}
  \label{fig:AARFFullZLE100}
\end{figure}

Fig.~\ref{fig:AARFFullZLE100} shows the charge distribution from FWF data and from ZLE data. The latter charge distribution is cut off below \SI{2}{pC} due to the ZLE threshold. The distributions are consistent above the \SI{2}{pC} threshold. Also shown is the charge distribution for dark noise pulses identified in the afterpulsing analysis (Sect.~\ref{sect:ap}). The dark noise distribution does not match the LED light distribution in the multi-PE region because of the different probabilities of observing multi-PE pulses. In the LED light distribution, multi-PE pulses are observed from the light source with probabilities following a Poisson distribution. The multi-PE pulses in the dark noise SPE distribution are attributed to real light pulses from scintillation in the detector and from Cherenkov events in the PMT glass and light guide acrylic. 

\subsection{The SPE charge distribution model} 
The model of the SPE charge distribution has three components: The first component, due to normal multiplication of electrons incident on the first dynode, is a Polya distribution. The Polya distribution results when events follow a sequence of Poisson processes proceeding with slightly different rate parameters \cite{Coates:1970di}, as is the case for electrons undergoing multi-step multiplication on the dynodes within a PMT. For large numbers of multiplied electrons, the Polya distribution approaches a Gamma distribution, which is used here:
\begin{equation}
\text{Gamma}(q; \mu,b) = \frac{1}{b\mu\,\Gamma(\frac{1}{b})} \Big(\frac{q}{b\mu}\Big)^{\frac{1}{b}-1} e^{-\frac{q}{b\mu}} \label{eq:gamma}
\end{equation}
where $\mu$ is the distribution mean and $b$ is an arbitrary shape parameter.

As shown to work in Ref.~\cite{Caldwell:2013oea} for a similar PMT type, the second model component is another Gamma distribution, empirically chosen to model incomplete electron multiplication, such as when the primary photoelectron impacts the second instead of the first dynode.
Following Ref.~\cite{Kaether:2012bm}, we add the third model component, describing scattering of the photoelectron on the dynode structure, as an exponentially falling term which is cut off at the mean value of the primary Gamma distribution. 

\begin{figure}[htbp] 
\centering
     \includegraphics[width=0.8\textwidth]{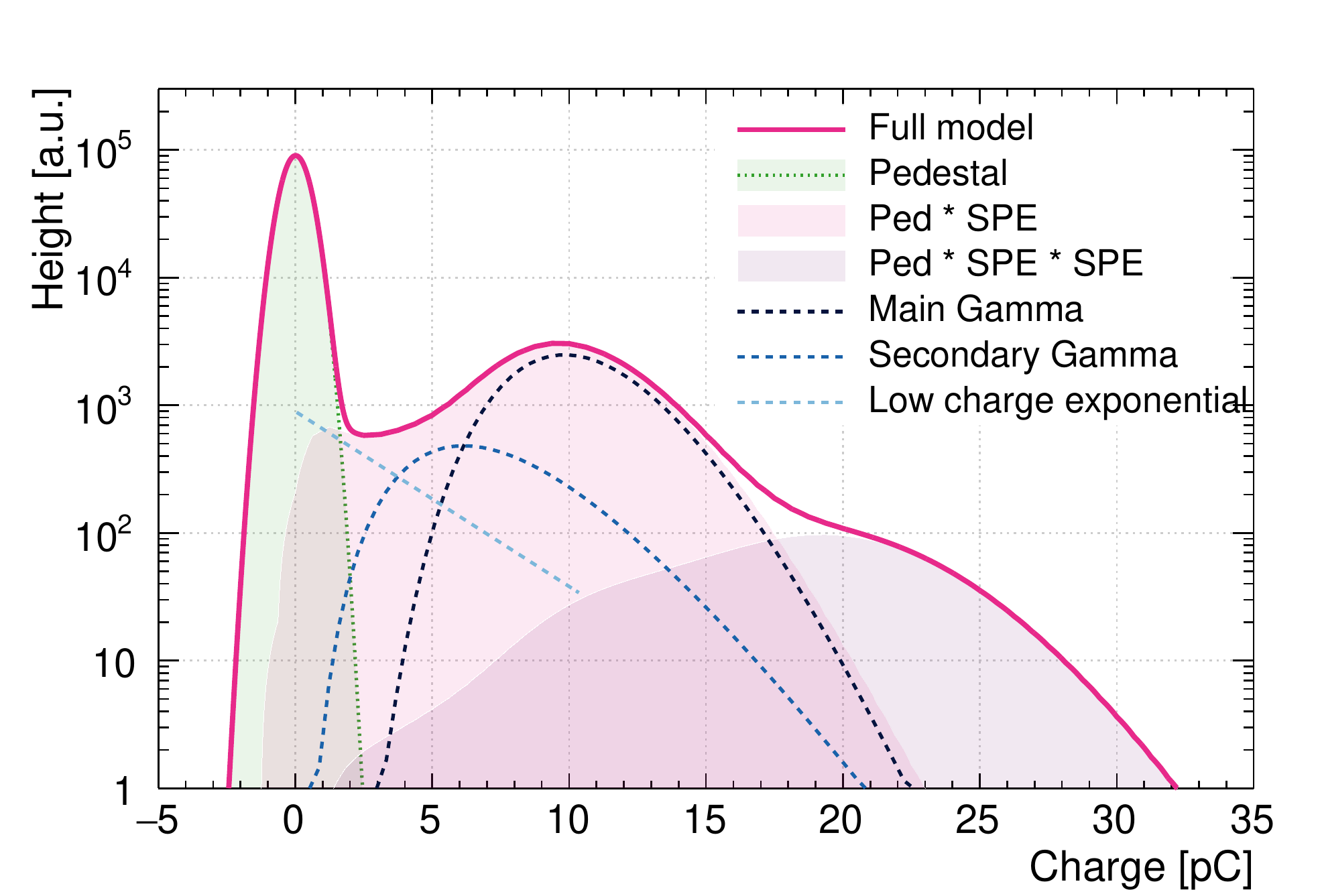}
     \caption[]{The thick solid line is the full low-light charge distribution model (Eq.~\eqref{eq:spemodel}) with parameters taken from a typical fit to a measured charge distribution. The pedestal, SPE, and 2-PE contributions to the full model are shown as solid filled curves. The SPE contribution is already convolved with the pedestal, and the 2-PE contributions is the convolution of the SPE contribution with itself and then with the pedestal. The SPE-distribution is further composed of three separate functions, drawn in dashed lines (before convolution with the pedestal).}
     \label{fig:spemodel}
\end{figure}

The full model is:
\begin{equation}
\text{SPE}(q) = \eta_1 \text{Gamma}(q; \mu,b) + \eta_2 \text{Gamma}(q; \mu f_{\mu}, b f_b) + 
\begin{cases} \eta_3 l e^{-q l} \label{eq:spe} \;(q < \mu) \\
0 \; (q > \mu)
\end{cases}
\end{equation}
The $\eta$ parameters are the amplitudes of each component and the entire distribution is normalized to unity. The three model components are shown graphically in Fig.~\ref{fig:spemodel}.
$f_{\mu}$ and $f_b$ are the relative position of the means and relative widths of the two Gamma distributions, so that adjusting the main Gamma distribution's parameters will also adjust the secondary distribution's parameters. In particular, changing the mean of the main Gamma distribution, which is the dominant influence on the mean-SPE charge, will cause the shape of the other two distributions to be adjusted as well. This is a crucial feature of the model function to enable weekly SPE calibration of all 255 PMTs without much human oversight and without the need to record the full SPE spectrum every week, as will be discussed in the next section.

\subsection{Fitting the low-light charge distribution} 
The measured low-light charge distribution includes a pedestal (zero-PE peak) and multi-PE contributions where the relative number of events from each contribution is determined by the occupancy through Poisson statistics. 
The width and shape of the pedestal is determined by the details of the electronics chain and is not a feature of the PMTs. To determine the shape of the pedestal, we created a charge histogram from data where the PMT bias voltage was turned off. We found that our pedestal is modelled well by a Gaussian distribution. 

\begin{align}
\text{Ped}(q) &= \frac{1}{\sqrt{2\pi}\sigma_\text{ped}} \text{exp}\left[-\frac{(q - \mu_\text{ped})^2}{2\sigma_\text{ped}^2}\right]
\end{align}

Each N-PE peak is modelled by the convolution ($\otimes$) of the SPE distribution N times with itself and once with the pedestal\cite{Wright:2007ju}. The full fit function is:

\begin{align}
\begin{split}
f(q) = & B\cdot\big[\;A\cdot \text{Ped}(q) + \text{Poisson}(1,\lambda) \cdot  \text{Ped}(q) \otimes \text{SPE}(q) \\
       &+ \text{Poisson}(2,\lambda) \cdot  \text{Ped}(q) \otimes  \text{SPE}(q) \otimes \text{SPE}(q) \\
        &+ \text{Poisson}(3,\lambda) \cdot  \text{Ped}(q) \otimes \text{SPE}(q) \otimes \text{SPE}(q)\otimes \text{SPE}(q) \\
& \; + ...\big] \label{eq:spemodel}
\end{split}
\end{align}
where $\lambda$ is the mean number of PE per flash of the optical source (approximately related to the occupancy as Eq.~\eqref{eq:gammaocc}) and $A$ and $B$ are arbitrary amplitudes. In principle, $A = P(0,\lambda)$, but to simplify calculating the charge distributions without the pedestal, we leave the pedestal height as an independent parameter so that the pedestal can be removed from the model (by setting $A = 0$) without affecting the N-PE distribution amplitudes. 

Fig.~\ref{fig:spemodel} shows the model from Eq.~\eqref{eq:spemodel} broken out into the 0-PE, SPE, and 2-PE components.

Eq.~\eqref{eq:spemodel} is fit to the FWF charge histograms for each PMT with all parameters allowed to float. 
The fit range is from \SI{-2}{pC} to $N_\text{max} \cdot$~\SI{10}{pC} where $N_\text{max}$ is chosen such that the ($N_\text{max}+1)$~PE peak has fewer than 1\% of pulses, based on the occupancy, and its contribution and that of any higher PE peaks is thus negligible.

FWF data is taken infrequently due to the large data volume, and FWF fits with all ten parameters floating are time intensive and sensitive to the choice of start parameters. 
For regular PMT gain monitoring, ZLE data sets are fit from \SI{+2}{pC} to $N_\text{max} \cdot$~\SI{10}{pC} while fixing all parameters to their fit values from the FWF fits, except $\mu$, the overall amplitude $B$, and the Poisson parameter $\lambda$ (Eq.~\ref{eq:gammaocc} is used to set the starting value for $\lambda$ but threshold effects cause the fitted value to be larger than the value calculated from Eq.~\ref{eq:gammaocc}). The peak position $\mu$ of the main Gamma distribution dominates the overall mean of the SPE distribution, and the secondary Gamma and exponential distributions' parameters are defined relative to $\mu$. Therefore, this constrained fit reproduces the SPE distribution shape and mean under the assumption that the SPE distribution stretches with the gain, but that the relative amplitudes of the three components do not change significantly on a time scale of months.

\begin{figure}[htbp] 
\centering
      \includegraphics[width=0.8\textwidth]{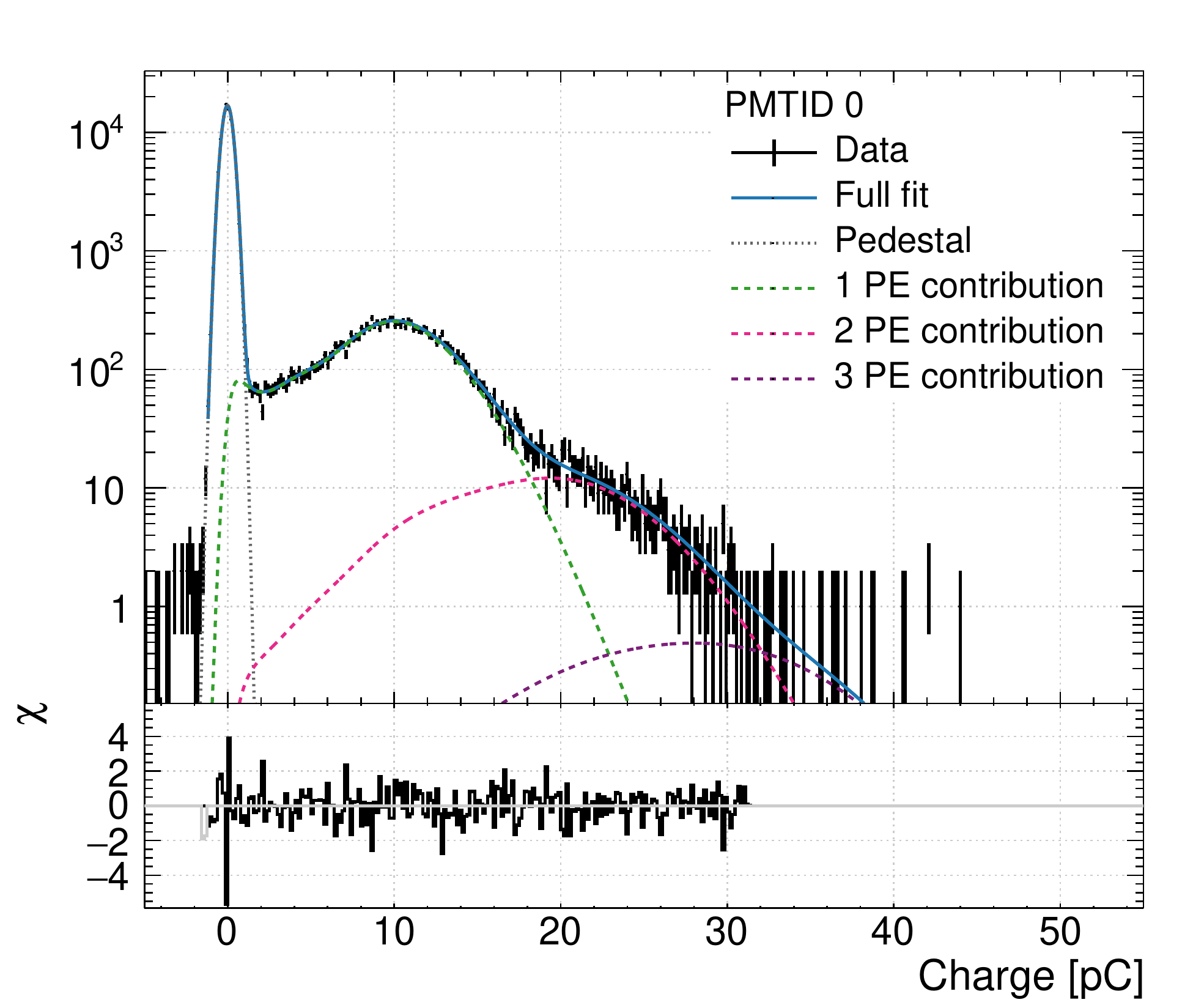}
      \caption[]{The low light charge distribution from FWF data for PMT~0 at 15\% occupancy is shown together with the model fit. The components of the fit function that represent the pedestal, 1, 2, and 3-PE charge distribution are also shown individually (compare Eq.~\eqref{eq:spemodel}). While most calibration data is taken at 5\% occupancy, this higher occupancy fit is shown to better illustrate the model description of the multi-PE components in the charge distribution. The conversion factor from charge to PE is 9.6~pC/PE based on this fit.}
\label{fig:highoccspe}
\end{figure}

\subsection{Evaluation of the mean SPE charge} \label{sec:meanspe} 

The mean charge of the SPE distribution is not  directly the fit parameter $\mu$; it is obtained from the model fit by taking the SPE contribution separately and calculating its mean (Eq.~\eqref{eq:spe}). We call this the \emph{fit mean SPE charge}. The SPE distribution's mean is always at a smaller charge than its maximum. 

For a simple model-independent consistency check similar to \cite{Saldanha:2017eq}, we calculate the  \emph{corrected histogram mean} $\hat{\mu}_{\text{SPE}}$; this is the mean of the measured low-light charge distribution histogram evaluated above the \SI{2}{pC} ZLE threshold ($\hat{\mu}_{\text{hist}}$), corrected for the multi-PE components using Poisson statistics. To calculate the Poisson correction, consider that neglecting threshold effects
\begin{align}
\begin{split}
\hat{\mu}_{\text{hist}} &= \hat{\mu}_{\text{SPE}} \frac{\text{P}(N=1,\lambda)}{\text{P}(N>0,\lambda)} + 2\cdot \hat{\mu}_{\text{SPE}} \frac{\text{P}(N=2,\lambda)}{\text{P}(N>0,\lambda)} + 3\cdot \hat{\mu}_{\text{SPE}} \frac{\text{P}(N=3,\lambda)}{\text{P}(N>0,\lambda)} + ... \\
   &= \frac{\hat{\mu}_{\text{SPE}}}{\text{P}(N>0,\lambda)} \sum_{i = 1}^{\infty} i P(N=i, \lambda) 
\label{eq:meanhist}   \end{split}
\end{align}
where $P(N, \lambda)$ is the Poisson function from Eq.~\eqref{eq:poisson}. Evaluating the sum and considering $\text{P}(N>0,\lambda) = 1-e^{-\lambda}$, we find that
\begin{equation} 
\hat{\mu}_{\text{hist}}    =  \frac{\hat{\mu}_{\text{SPE}}}{1-e^{-\lambda}} \cdot \lambda
\end{equation}
and using Eq.~\eqref{eq:gammaocc} it follows that the actual mean of the SPE charge goes with the histogram mean and the occupancy as
\begin{equation}
\hat{\mu}_{\text{SPE}} \simeq \frac{\hat{\mu}_{\text{hist}} \cdot \mathcal O}{-\text{ln}(1 - \mathcal O)} \label{eq:occcorrection}
\end{equation}
where the `$\simeq$' indicates that this is an approximate relation due to threshold effects.

Since the histogram mean can only be obtained with the \SI{2}{pC} threshold imposed and is thus not directly comparable to the fit mean, we additionally calculate the \emph{fit mean above \SI{2}{pC}}.

\begin{figure}[htbp] 
     \includegraphics[width=0.8\textwidth]{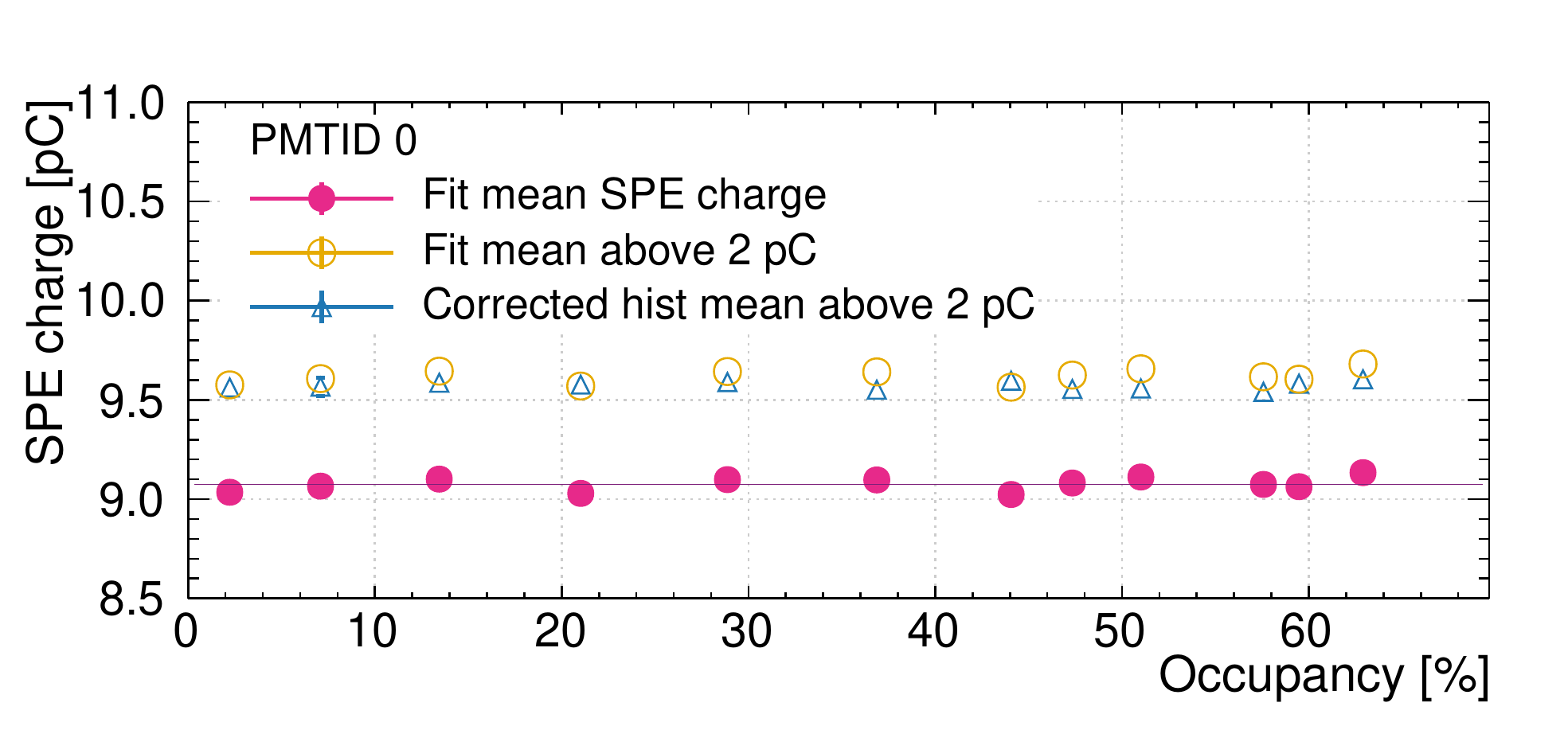}
     \centering
     \caption{SPE charge from ZLE data. The mean of the SPE-component of the low-light charge distribution fit, the estimated SPE mean from the occupancy corrected mean of the low-light charge histogram above \SI{2}{pC}, and the mean of the fit function above \SI{2}{pC} are shown for different occupancies. Statistical error bars are smaller than the marker size. }
     \label{fig:spe_by_occ}
\end{figure}

Fits to FWF data for a representative PMT with 15\% occupancy are shown in Fig.~\ref{fig:highoccspe}. The fit mean SPE charge is shown in Fig.~\ref{fig:spe_by_occ} as a function of occupancy for the same PMT. It is stable within uncertainties over a range of occupancies from 3\% to 60\%, with $\chi^2$/ndf between 0.7 and 1.5.

The corrected histogram mean and the fit mean above \SI{2}{pC} are also shown in Fig.~\ref{fig:spe_by_occ}. The histogram mean is higher than the fit mean because it is biased by the threshold; it is consistent within uncertainties with the fit mean above \SI{2}{pC}.

The largest uncertainty in the SPE charge distribution comes from the low-charge exponential, which is dominant only in a small section of the charge spectrum, and otherwise obscured by the pedestal.
We determine the mean SPE charge at extreme values of the low-charge exponential to demonstrage to what extend the details of this term affect the result: Without a low-charge exponential, the mean of the SPE charge distribution would be approximately 9\% higher. If the low-charge exponential had twice the amplitude of the typical value observed, the SPE distribution mean would be 6\% percent lower. 

\begin{figure}[htbp] 
     \includegraphics[width=0.8\textwidth]{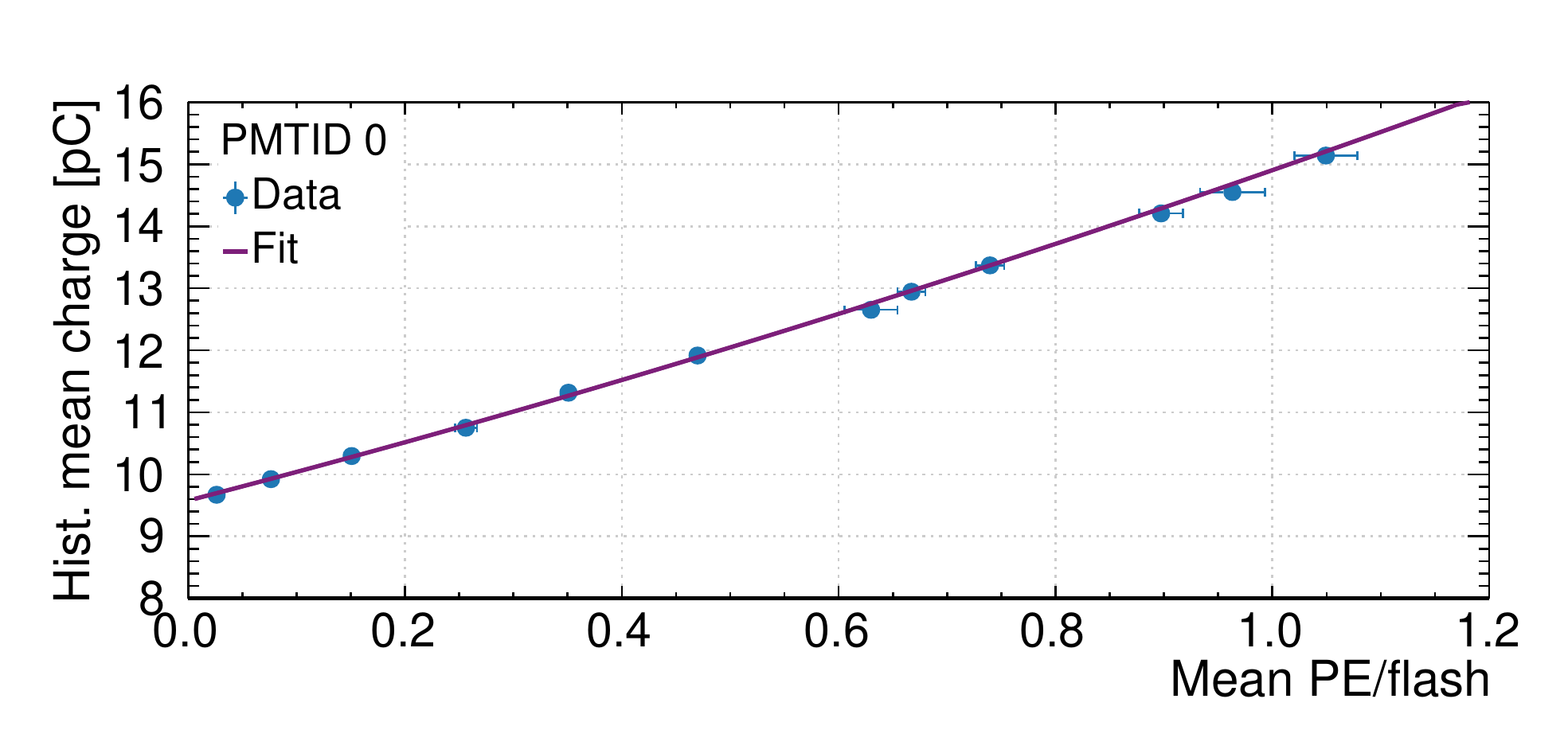}
     \centering
     \caption{The effect of the ZLE threshold is shown. The mean charge of the low-light charge histogram is plotted as a function of the measured mean charge per optical flash for the given run. The fit gives a mean SPE charge of $9.33\pm 0.06$ pC. The threshold effect ($\delta$ in the text) is  $2.6\pm 0.9$\% of mean SPE charge.}
     \label{fig:spe_by_occ_fit}
\end{figure}

To remain compatible with the observed distributions, the parameters of the exponential cannot take on these extreme values. We use Eq.~\ref{eq:meanhist} to constrain the below-threshold effect accurately. The threshold affects the mean charge of SPE signals but has a negligible effect on multi-PE signals. Mathematically, the mean charge measured for an n~PE pulse will be $n\cdot \hat{\mu}_{\text{SPE}}$ while for SPE pulses it will be $(1-\delta)\hat{\mu}_{\text{SPE}}$, where $\delta$ is the correction in charge imposed by the threshold. We now plot the mean number of PE per flash (determined from Poisson statistics) against the mean of the low-light charge histogram for data used in Fig.~\ref{fig:spe_by_occ}. This is shown in Fig.~\ref{fig:spe_by_occ_fit}. The data is fit with Eq.~\ref{eq:meanhist} after including the $\delta$-term. Based on the value of $\delta$ obtained in the fit, we assign a systematic uncertainty on the mean SPE charge of 3\%. This is comparable to other measurements \cite{Anthony:2018jv}.

This calibration is performed for each PMT in the detector. We find that the charge spectra are similar across the PMT array. The average values and RMS (in brackets as percentage of the average) of the relevant fit parameters are the following: $f_{\mu}$=0.61 (7\%), $f_b$= 3.2 (15\%), $\eta_2$=0.26 (16\%), $\eta_3$=0.16 (25\%), $\chi^2/\text{NDF}$=0.99 (17\%); the largest value we obtain for the reduced chi-squared in a fit is $\chi^2/\text{NDF}$=1.49.


\subsection{Mean SPE charge dependence on bias voltage} \label{sect:charge_biasv} 

The mean SPE charge $\overline{q}$ depends on the bias voltage U as
\begin{equation}
\overline{q} = \text{A}\cdot \text{U}^\gamma \label{eq:gainvsbias}
\end{equation}
where A is an arbitrary constant. Knowledge of the $\gamma$ parameter for a PMT is necessary to quantify the effect of fluctuations in the bias voltage and to adjust the mean SPE charge in order to match the gains across the PMT array.

\begin{figure}[htbp] 
\centering
     \includegraphics[width=0.8\textwidth]{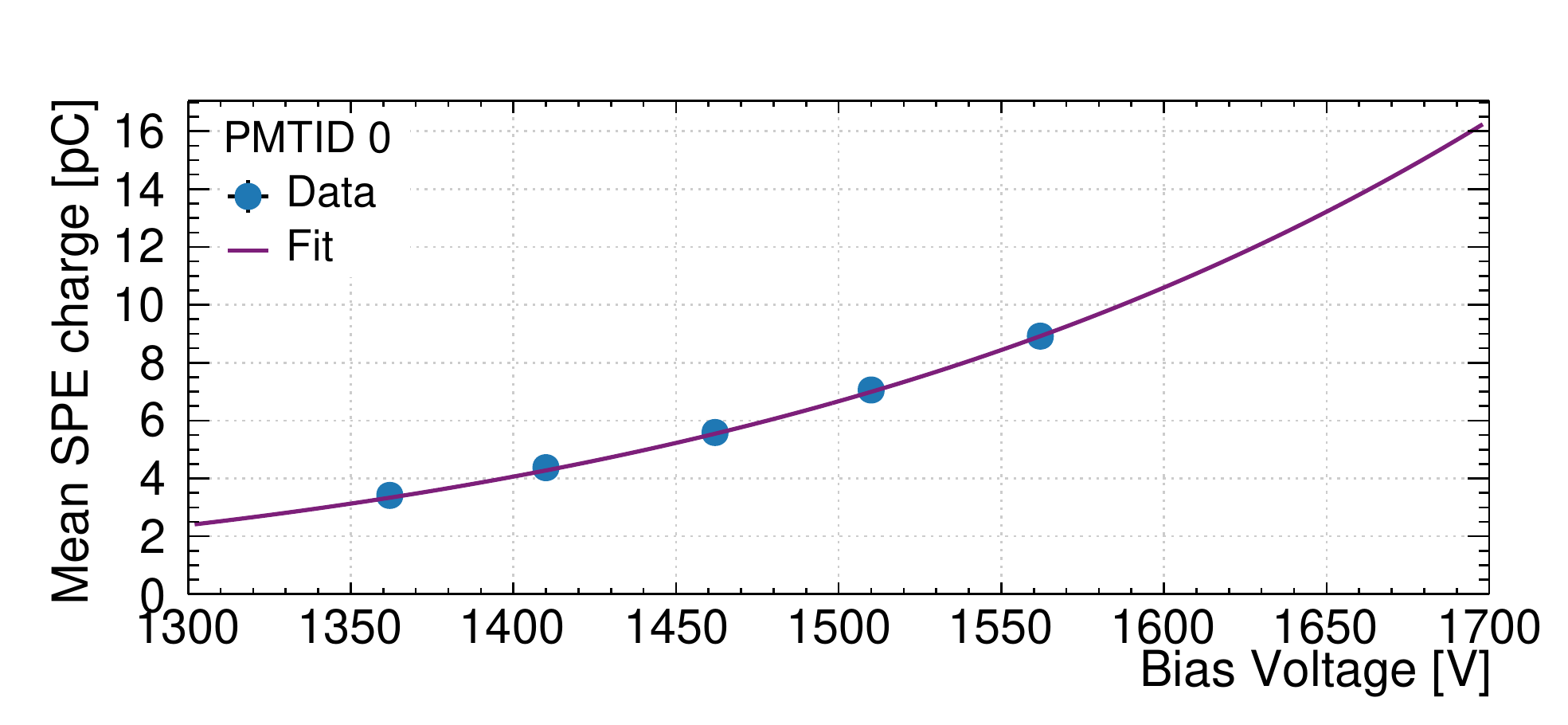}
     \caption[]{Bias voltage versus fit mean SPE charge for PMT~0. This PMTs nominal bias voltage is \SI{1560}{V}. The fit result is $\gamma = 6.97 \pm0.01$ at $\chi^2/$NDF = 2.4/3.}
     \label{fig:gainvsbias}
\end{figure}

LED calibration data were taken with the PMTs at \SI{-200}{V}, \SI{-150}{V}, \SI{-100}{V}, \SI{-50}{V}, and \SI{0}{V} relative to their nominal bias voltage. The fit mean SPE charge for each dataset versus bias voltage for a sample PMT is shown in Fig.~\ref{fig:gainvsbias}. Eq.~\eqref{eq:gainvsbias} is fit to the points in order to obtain the $\gamma$ parameter. Note that the actual voltage on the PMT is approximately \SI{6}{\%} lower than the supplied voltage due to a series resistor between the PMT cable and the high voltage supply.

\subsection{Discussion}

For the Hamamatsu R5912-HQE, we developed an effective three component model of the SPE charge distribution which describes normal photoelectron multiplication as well as nuisance effects from electron scattering and incomplete multiplication. 
The model SPE distribution is integrated in a sum model to fit the complete measured low-light charge histogram, and extracted again after the fit to determine the SPE distribution mean for a PMT.

The full model fits the data without systematic deviations in an occupancy range from under 3\% to over 60\%. Applying a method similar to that of~\cite{Saldanha:2017eq} successfully described the effect of pedestal and the charge response as light yield increased.


\section{Measuring double and late pulsing probabilities}\label{sect:double}
Photoelectrons produced at the photocathode can backscatter off the first dynode or the grid. Double pulsing refers to inelastic scattering, where a fraction of the photoelectron energy is deposited in the initial backscatter, while the rest goes with the backscattered electron. As the name suggests, the initial scatter causes a pulse in the PMT waveform that is then followed by a second pulse as the backscattered electron returns to the dynode chain. The sum of the integral charge of these two pulses will correspond to the characteristic single photoelectron charge of the PMT. Late pulsing refers to elastic backscattering off the first dynode. A pulse in the PMT waveform does not occur until the scattered electron returns to the dynode, producing a broadened response time.  These effects occur on a time-scale less than \SI{100}{ns} after the light flash. Correlated noise at times scales $> 100$~ns is discussed in Sect.~\ref{sect:ap}.

Double and late pulsing charge and time probability distributions were measured using the \SI{445}{nm} laserball data and DAQ setting 2 from Tab.~\ref{tab:daqsettings}. 

The procedure to characterize double pulsing in the PMTs is described in Sect.~\ref{sec:dp}. This will inform the late pulsing characterization outlined in Sect.~\ref{sec:lp}.

\subsection{Double pulsing} \label{sec:dp}
A pulse is called a \emph{primary pulse} if it occurs in a strict time window near the light injection time. This acceptance window is defined as follows. We create the pulse time distribution for the given PMT (see Fig.~\ref{fig:gammafit}). We set the lower edge of the window to include the full rise time to the left of the peak; in Fig.~\ref{fig:gammafit} this is at approximately \SI{-6}{ns}. The upper edge is set at the point when the pulse time distribution drops to 5\% of the peak height (in Fig.~\ref{fig:gammafit} this is at approximately \SI{24}{ns}). The window is typically \SI{30}{ns} wide.

 A double pulse is recorded if a primary pulse exists and if the subsequent pulse (called the \emph{follower pulse}) occurs within \SI{100}{ns}. To minimize contamination by afterpulses from earlier dark noise pulses, events are discarded if they have pulses in the \SI{6.5}{$\mu$s} before the acceptance window.

\begin{figure}[htbp]
\centering
\includegraphics[width=0.8\textwidth]{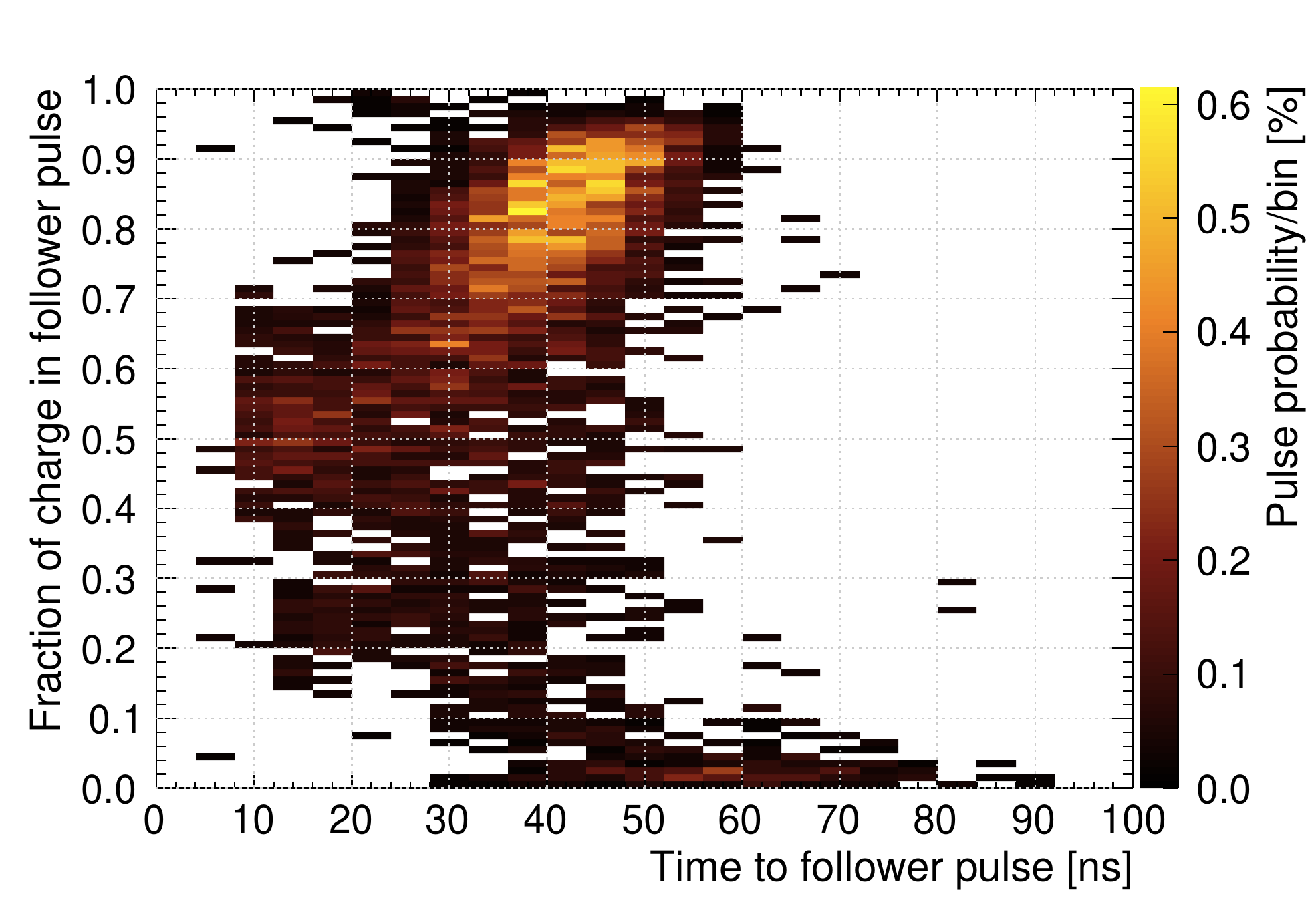}
\caption{Fraction of charge in the follower pulse over the total charge of lead and follower pulse versus the arrival time after the lead pulse for PMTID~0. The dark noise component has been subtracted (see text) and negative bins are suppressed.}\label{fig:doubleqt}
\end{figure}

\begin{figure}[h!]
\centering
\includegraphics[width=0.8\textwidth]{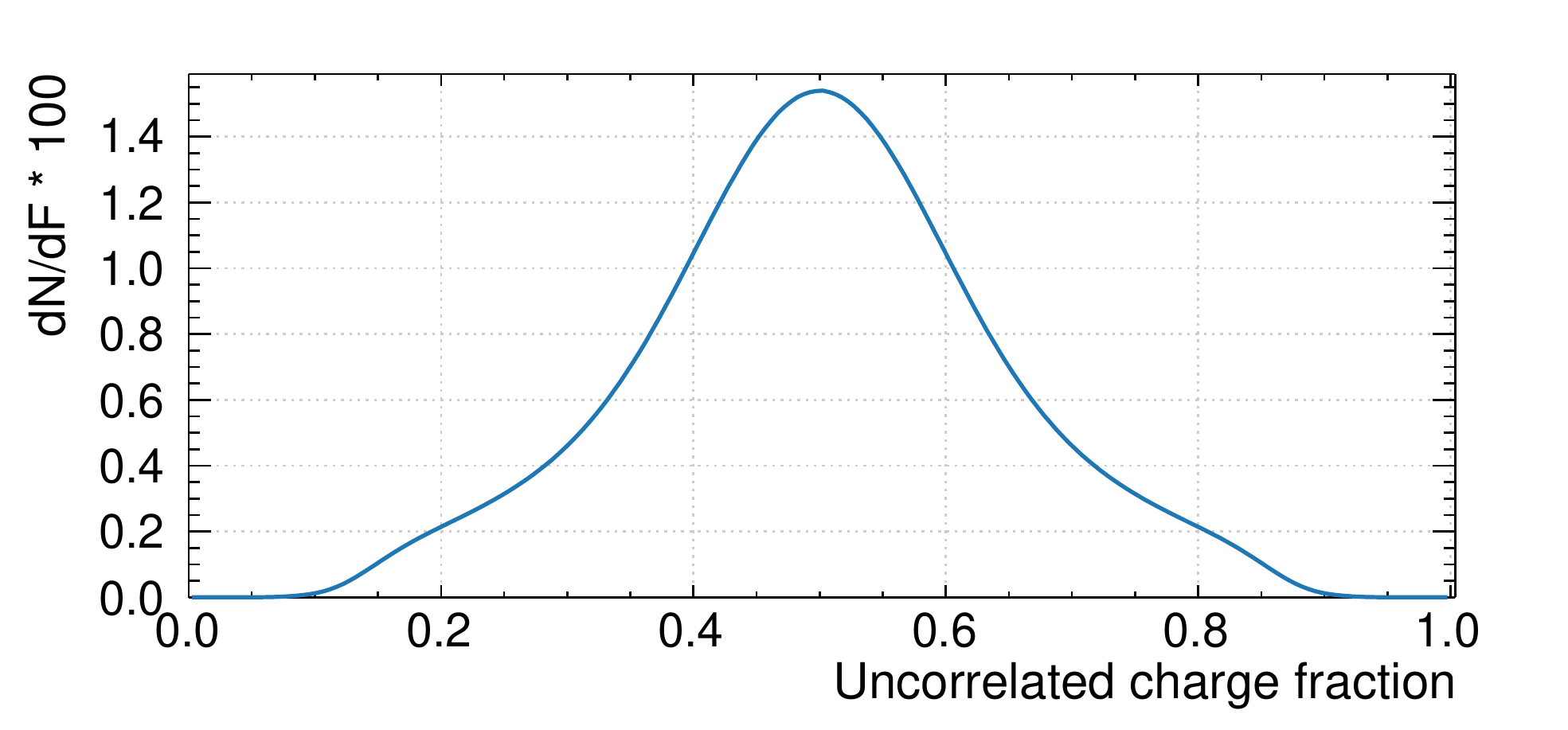}
\caption{Charge fraction distribution for two uncorrelated pulses in PMT 0. This distribution is used to subtract the dark noise component of the double pulse joint charge time PDF.}\label{fig:uncorrcharge}
\end{figure}

The pulse times and charges were recorded for events which matched these criteria, and the fraction of charge in the follower pulse relative to the sum charge of the primary and follower pulse, and the follower pulse arrival time, were binned into a histogram. This histogram was normalized to unity to create a probability density function (PDF). An example of such a joint charge and time PDF is shown in Fig.~\ref{fig:doubleqt}. The dark noise component was subtracted from the distribution as follows. Dark noise pulses have a uniform time distribution in the charge-time distribution, as they are uncorrelated with the primary pulse. They also have a characteristic charge fraction distribution, again, as the two pulses are uncorrelated. This distribution, illustrated in Figure \ref{fig:uncorrcharge}, is symmetrical around 0.5 with the shape determined by the SPE charge distribution of the PMT in question. Once normalized to the expected dark rate per time bin, the dark noise distribution was subtracted from the double pulse charge fraction, leaving only the double pulsing joint PDF shown in Fig.~\ref{fig:doubleqt}. 

\begin{figure}[h!]
\centering
\includegraphics[width=0.6\textwidth]{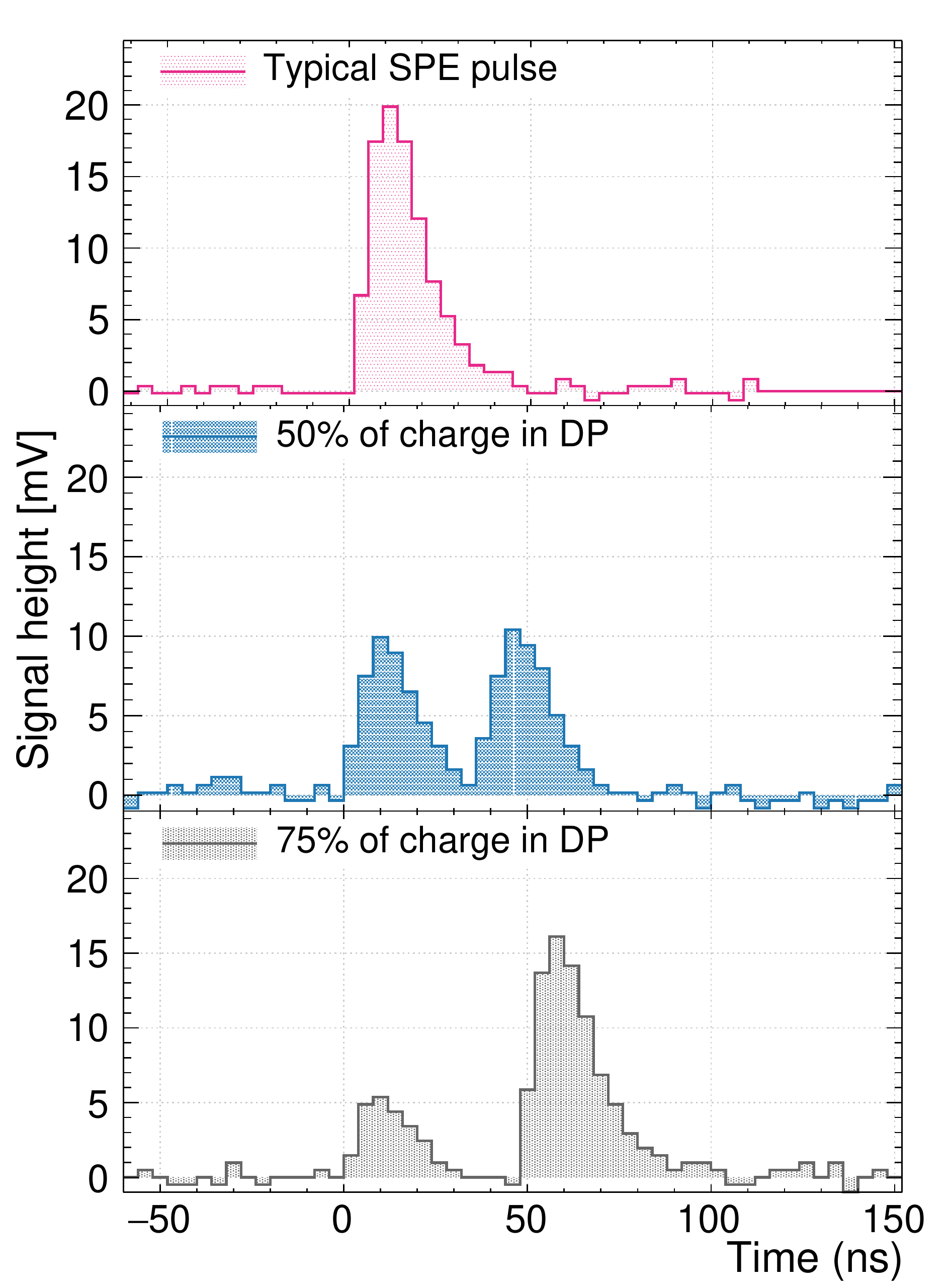}
\caption{Three representative PMT waveforms resulting from a single PE released at the photocathode are shown. The voltage is sampled, not integrated, in the \SI{4}{ns} sampling window. The raw data has a positive voltage baseline and the signal corresponds to a drop in voltage. The raw data was inverted and the baseline subtracted. The top panel shows a typical SPE-size pulse. The middle and bottom panels show double pulses where the follower pulse contains 75.2\% and 50.1\% of the total charge.}
\label{fig:qfracBoth}
\end{figure}

Distinct regions can be seen in Fig.~\ref{fig:doubleqt}, the most prominent of which has a charge fraction centered around 0.85 and arrival time of approximately \SI{45}{ns}. This arrival time corresponds to approximately twice the photoelectron transit time from the photocathode to the dynode. Fig.~\ref{fig:qfracBoth} shows examples of double pulse waveforms with charge fractions of 0.75 and 0.5 respectively.

The double pulse probability, $p_{DB}$, is the probability that a primary pulse will cause a double pulse. It is given by

\begin{equation}
  p_{\rm{DB}} = \frac{n_{\text{DB}}}{N_{\text{Init}}}
\end{equation}
where $n_{\text{DB}}$ is the number of double pulses and $N_{\rm{Init}}$ is the number of primary pulses recorded.

To calculate $n_{\text{DB}}$ one must account for the fact that a double pulse is only recorded when it is not preceded by a dark noise pulse or a second laser pulse (from scattered light). The double pulse distribution must, therefore, be {\em unshadowed} where the number of events recorded at a particular time is adjusted according to the probability of events occurring at earlier times. This method is described in detail in \cite{retiere:2017}. After unshadowing, the number of dark pulses and the number of times when the PMT observed more than one photon from the laser flash must be subtracted. The number of dark pulses, $n_{D}$, is estimated using Poisson statistics where:
\begin{equation}
  n_{\rm{D}} = {\text{Poisson}}(1,R_{\text{DN}}T)N_{\text{Init}}.
\end{equation}
Here the expected number of dark pulses is given by the measured dark rate for the PMT, $R_{\text{DN}}$, multiplied by the length of the time acceptance window $T$. The Poisson probability of seeing one dark pulse in this window is then multiplied by the number of initial pulses. This gives the total number of dark pulses recorded as follower pulses. The number of secondary laserball pulses, $n_{\rm{SL}}$, is calculated using the Poisson probability given the occupancy $\mathcal{O}$. Here we define the occupancy as the expected number of laserball pulses, $n_{\text{LB}}$, in the PMT per event. So we have
\begin{equation}
n_{\rm{SL}} = {\text{Poisson}}(n_{\text{LB}} > 1,{\mathcal{O}})N_{\text{Events}},
\end{equation}
where the probability is multiplied by the total number of recorded events $N_{\rm{Events}}$. The expected number of double pulse events is then given by 
\begin{equation}
  n_{\text{DB}} = n_{\text{Tot}}-n_{\text{D}}-n_{\rm{SL}}.
\end{equation}\label{eqn:pdb}

Here, $n_{\text{Tot}}$ is the total number of follower pulses recorded. 

A typical DEAP PMT has a double pulsing probability of $p_{\rm{DB}} = 2.7\%\ \pm 0.01 \%$.

\subsection{Late pulsing} \label{sec:lp}
\begin{figure}[htbp]
\centering
\includegraphics[width=0.8\textwidth]{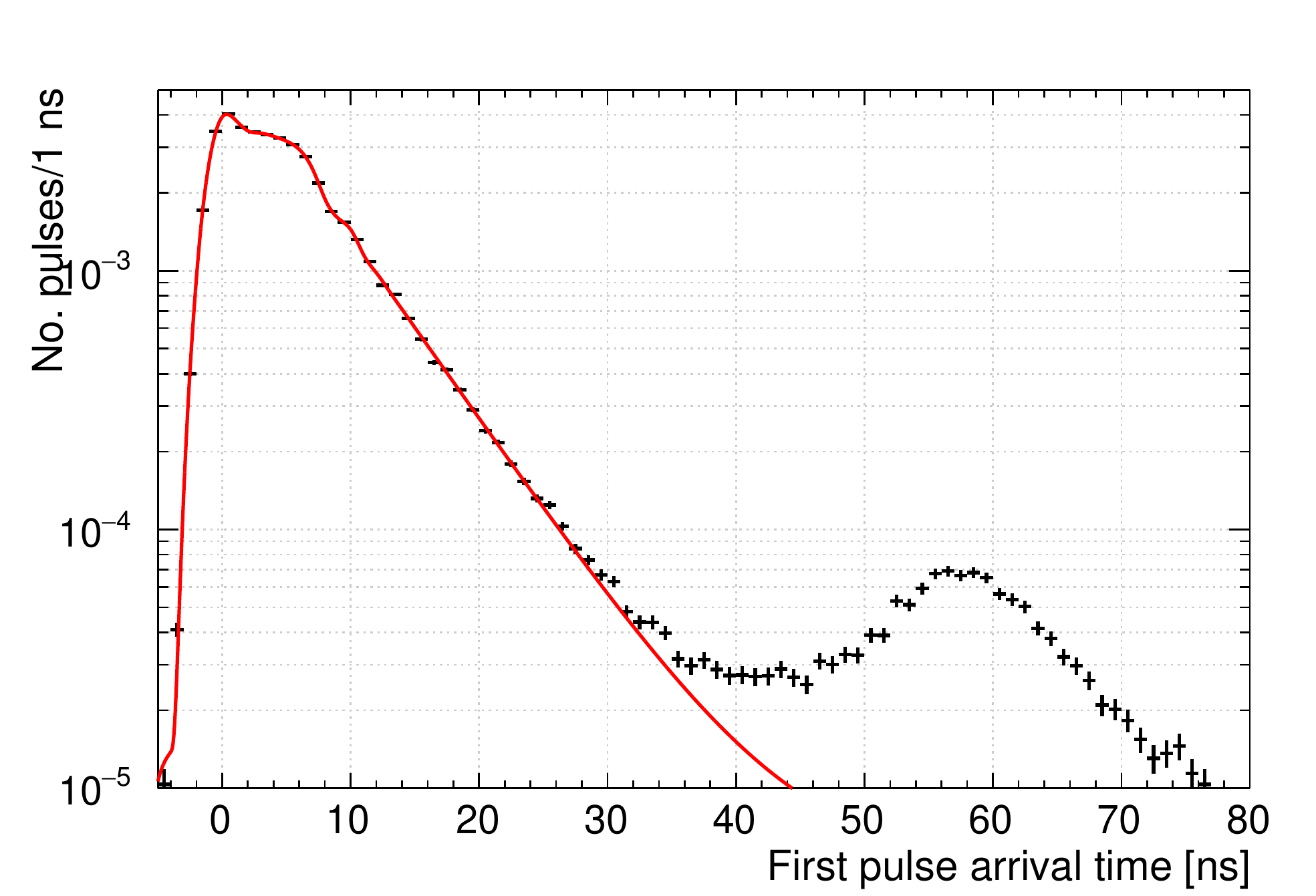}
\caption{Distribution of first photon arrival times. The primary pulse time distribution is fit using a piecewise combination of a flat dark rate component, an exponential function, and an Akima spline. This determines the extent of the primary pulse time distribution and upon subtraction leaves only the late pulse distribution in the recorded data. Late pulses can be seen between 35 and 80 ns. The observed late pulse distribution is a convolution between the optical detector response function and the true late pulse time PDF. Deconvolution is not possible from an in-situ measurement.} \label{fig:gammafit}
\end{figure}

Late pulsing involves the absence of an initial pulse at the time the laserball fired since the photoelectron elastically scatters off the dynode stack. In order to study the time distribution of late pulses, events were selected where the first pulse arrived at any time after the start of the primary pulse acceptance window as described above. The time of this first pulse in each PMT waveform was recorded; an example of this distribution is shown in Fig.~\ref{fig:gammafit}. 

The measured distribution is the sum of the arrival time distributions of primary pulses, late pulses, and dark noise pulses. In order to obtain the distribution of late pulses by itself, a model of the time distribution of primary pulses must be found, so that this component may be subtracted from the measured distribution. Reflections within the detector, and therefore the positions of the PMTs, are the dominant factor determining the shape of the peak (a photon needs approximately \SI{6}{ns} to cross the inner detector vessel). The region relevant to the analysis is the late time ($>$ \SI{25}{ns}) tail of the distribution which overlaps with the late pulsing region. This can be modeled as a Poisson process using an exponential function. A first-principles optical model of the detector response has been developed but is beyond the scope of this paper. The empirical model used here is a piecewise combination of an exponential function, a flat contribution, and an Akima spline, shown in Fig.~\ref{fig:gammafit} as a red line and described mathematically as:
\begin{equation}
f_{LB}(t) = 
\begin{cases}
A(t) & \quad \text{if } t < t_{25}  \\
\frac{B}{\tau}\exp\left( -\frac{t}{\tau} \right)+c & \quad \text{if } t  \geq t_{25}.\\
\end{cases}
\label{eq:primarypulsetime}
\end{equation}
$t_{25}$ represents the time at which the primary pulse peak (near t=0~ns in Fig.~\ref{fig:gammafit}) has fallen to 25\% of its maximum. The Akima spline $A(t)$ describes this region, which has no effect near the late pulsing peak at \SI{58}{ns} and is therefore sufficiently modelled by the spline. In the late pulsing region, the primary pulse distribution becomes an exponential function with normalisation $B$ and time constant $\tau$ above the flat dark rate component $c$. The parameters $B$, $\tau$, and $c$ are determined through fitting. The exponential tail and dark rate component fit well across all PMTs while, as mentioned above, the shape of the peak of the distribution depends heavily on the position of the PMT relative to the laserball.

Upon subtraction of Eq.~\eqref{eq:primarypulsetime} from the measured time distribution, only the late pulse distribution remains. This distribution is, however, still convolved with the optical detector response.
Deconvolution would have to be performed to accurately determine the shape and extent of the true underlying late pulsing distribution. Even though this optical response function is given by Eq.~\eqref{eq:primarypulsetime}, in-situ measurements do not have the time resolution necessary for such a deconvolution to yield a smooth PDF. The late pulse distribution is expected to drop off sharply at twice the PMT transit time ($\sim$ \SI{56}{ns}) as shown in, for example, \cite{Brack:2012ig} and \cite{Kaether:2012bm}. This corresponds to the limiting case of an elastically scattered photoelectron traveling back to the photocathode before returning to the dynode.

The late pulse probability, $p_{\rm{LA}}$, can be determined directly from the late pulse distribution using the total number of late and dark pulses, $N_{\rm{LA}}$ and $N_{\rm{DN}}$ respectively. The number of late pulses given some total number of events, $N_{\rm{Events}}$, is given by
\begin{equation}
  N_{\rm{LA}} = p_{\rm{LA}} \cdot \mathcal O \cdot {\rm{Poisson}}(0;R_{\rm{DN}}T) \cdot N_{\rm{Events}}.
\end{equation}
Here the occupancy must be included in the calculation, for in order to get a double pulse, an unseen primary pulse must have occured. In addition, a dark noise pulse must not have occured which is given by ${\rm{Poisson}}(0;R_{\rm{DN}}T)$. In contrast, the number of dark pulses recorded depends on the primary pulse not appearing, so we have
\begin{equation}
  N_{\rm{DN}} = (1-\mathcal O)\cdot{\rm{Poisson}}(1;R_{\rm{DN}}T) \cdot N_{\rm{Events}}.
\end{equation}
Taking the ratio of these two values removes the dependence on the total number of events giving:
\begin{equation}
\frac{N_{\rm{LA}}}{N_{\rm{DN}}} = \frac{p_{\rm{LA}}\cdot \mathcal O \cdot {\rm{Poisson}}(0;R_{\rm{DN}}T)}{(1-\mathcal O)\cdot{\rm{Poisson}}(1;R_{\rm{DN}}T)},
\end{equation}\label{eqn:pla}
which can be rearranged to find the late pulsing probability. 
A typical DEAP PMT has a late pulsing probability of $p_{\rm{LA}} = 2.3\%\ \pm 0.05 \%$.

\section{Measuring afterpulsing probabilities}\label{sect:ap}
Afterpulses, a type of correlated noise pulses, are caused when a residual gas atom inside the PMT bulb becomes ionized by moving electrons. The positive ion is accelerated toward the photocathode or the first dynode, where it liberates new electrons that are multiplied and create a charge signal. Afterpulses occur from a few hundred nanoseconds to several microseconds after the initial pulse.

Afterpulsing randomly adds counts to certain areas of the waveform. 
In a PSD analysis, this makes the discrimination parameter distributions for both signal and background wider and moves them closer together \cite{butcher:2016}. In cases where a certain particle type is considered a background, this effect decreases the power of PSD to remove the background. In order to mitigate this effect, the afterpulsing characteristics need to be known for each PMT.

In this section, we study the time and charge distribution of afterpulses, as well as the overall afterpulsing rates, using the next-pulse charge and time distribution.

\subsection{Measurement of the next-pulse charge and time distribution}

Afterpulsing was studied using the LED calibration system with setting 3 from Tab.~\ref{tab:daqsettings}. The light intensity was such that the majority of PMTs had 5\% to 10\%
occupancy. A pulse in a given PMT that occurs within 80~ns of the LED flash is referred to as a \emph{primary pulse}. Individual pulses within this window are not separated, so double pulses (Sect.~\ref{sec:dp}) are counted as one pulse with the total integrated charge of both components. The window is large enough to contain late pulses (Sect.~\ref{sec:lp}) as well.

For each PMT, events were selected for which a primary pulse existed with no pulses in the \SI{20}{$\mu$s} prior to the light injection time. The charge of the primary pulse, and the time to and charge of the next pulse (the \emph{follower pulse}) were recorded. No further pulses were considered so as to exclude afterpulsing of afterpulses.

\begin{figure}[htbp] 
\centering
     \includegraphics[width=0.8\textwidth]{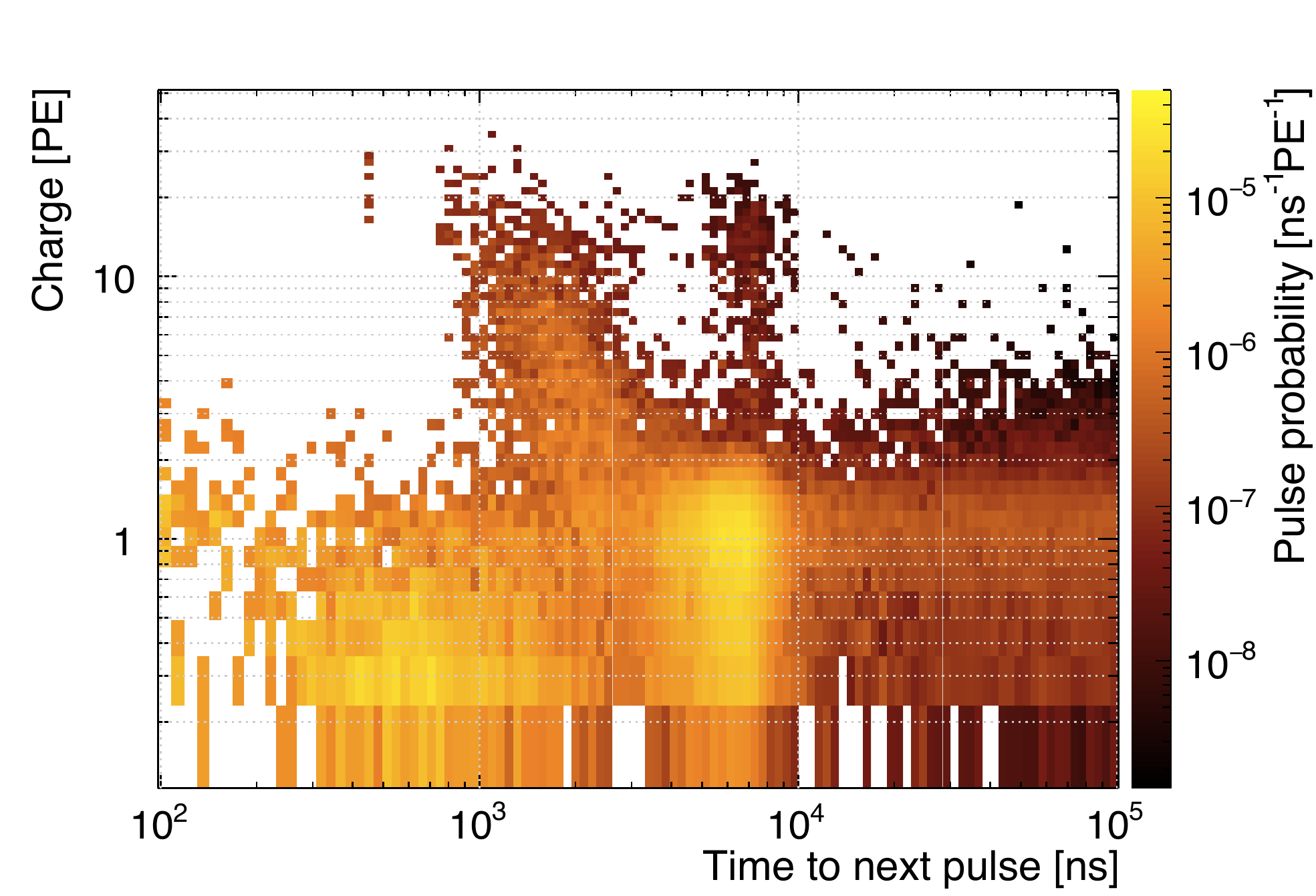}
     \caption[]{The measured probability for a typical PMT of observing a pulse following a primary pulse as a function of both
       the second pulse's charge (in units of the mean SPE
       charge) and of time. The primary pulses were required to have a charge between 10~pC and 14~pC in this example.}
     \label{fig:ap-hist-example}
\end{figure}

The charge and arrival time of the follower pulse are binned into a histogram. The bin widths of the x-axis increase exponentially. Bin contents and uncertainties
are normalised to the bin size to result in a probability per $\mbox{ns}\cdot\mbox{PE}$, where
PE is the measured charge in units of the mean SPE charge. Fig.~\ref{fig:ap-hist-example} shows this \emph{next-pulse charge and time} (NPCT) histogram for primary pulses of charge \SI{10}{pC} to \SI{14}{pC} and for a PMT with typical behaviour.

\subsection{Afterpulsing populations}
Several distinct event populations are visible in Fig.~\ref{fig:ap-hist-example}. To study them further, projections onto the time and charge axis are considered.

\begin{figure}[htbp] 
	\centering
     \includegraphics[width=0.8\textwidth]{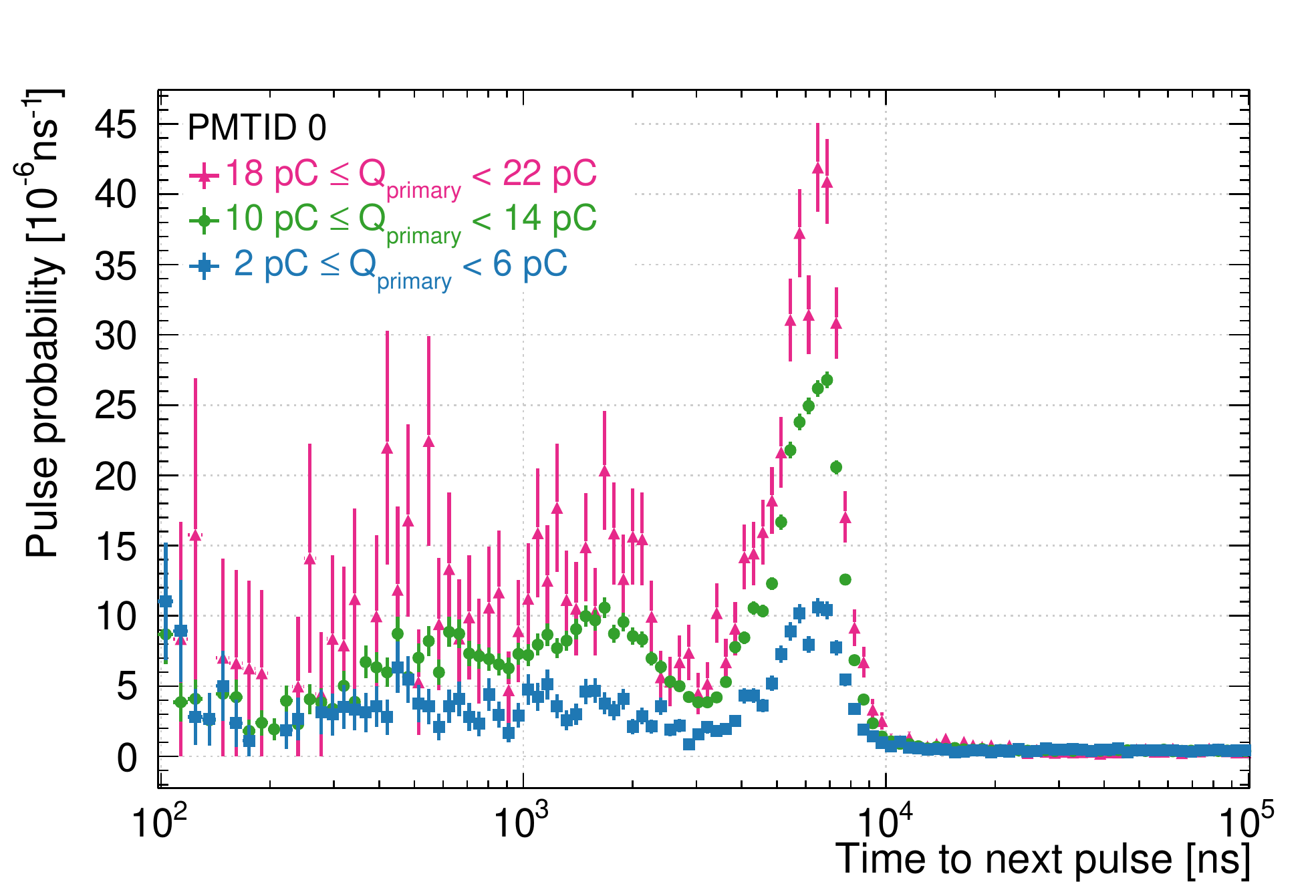}\\
     \caption[]{ The distribution of the time to next pulse
       is shown for an example PMT with typical behaviour for three ranges of primary pulse
       charge. SPE signals dominate in all charge bins. The data are not weighted for the charge of the follower pulse. The mean SPE charge for this PMT is 9.5~pC. The number of entries in each histogram varies strongly due to the shape of the SPE charge distribution, hence the statistical uncertainties shown are of different size between histograms.}
     \label{fig:ap-1d-t}
\end{figure}

The \emph{time to next pulse} (TTNP) distributions for primary pulses in three different charge bins are shown in Fig.~\ref{fig:ap-1d-t}. 
Afterpulse populations occur in three broad time ranges: \SI{200}{ns} - \SI{800}{ns}, \SI{800}{ns} - \SI{3000}{ns}, and \SI{3000}{ns} - \SI{10000}{ns}.
We attribute the smooth part of the distribution past \SI{10}{$\mu$s} to dark noise. 
\begin{figure}[htbp] 
	\centering
     \includegraphics[width=0.8\textwidth]{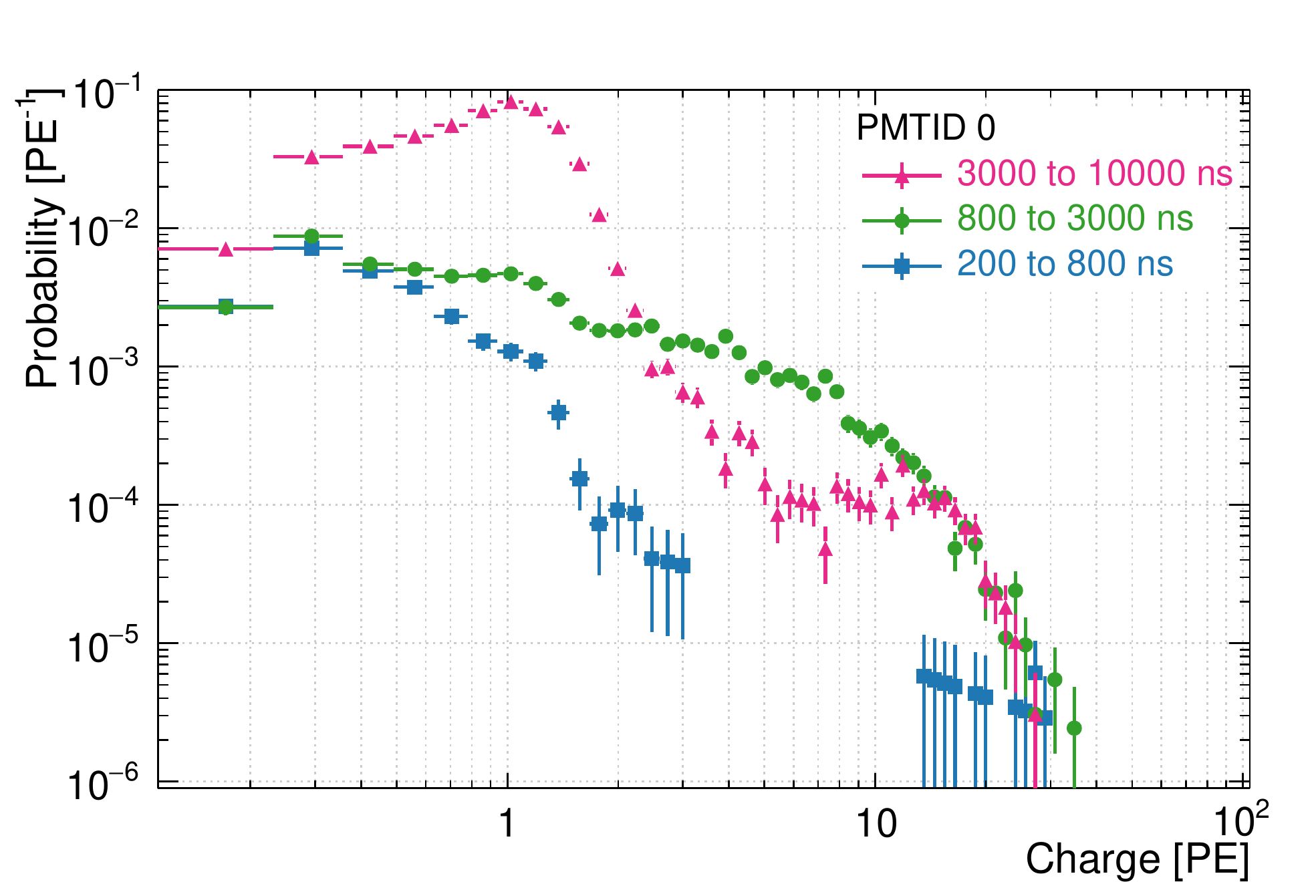}
     \caption[]{The charge distribution for afterpulses that occur in the three broad time ranges encompassing the peaks in Fig.~\ref{fig:ap-1d-t}.}
     \label{fig:ap-q-dist}
\end{figure}

The charge distributions of pulses arriving in the above three afterpulsing time ranges are shown in Fig.~\ref{fig:ap-q-dist} (the dark noise charge distribution is shown in Fig.~\ref{fig:AARFFullZLE100}). The average charge of afterpulses in each population is 0.6~PE, 2.1~PE, and 1.0~PE respectively. We observe afterpulses several 10's of PE in size.

We note that a small but distinct population of afterpulses occurs in a narrow time window between 420 to \SI{450}{ns} and at charges above 10 PE in Fig.~\ref{fig:ap-hist-example} (the poulation forms a narrow vertical band at the top left). This population is present in most of the 255 DEAP PMTs. Study of the waveforms in question indicates that these are real afterpulses and not an instrumental artefact.

\subsection{Total afterpulsing probability and its dependence on primary-pulse charge}
The total afterpulsing probability is the probability to have an afterpulse at any time and with any charge. 
This is calculated starting from the TTNP histograms (compare Fig.~\ref{fig:ap-1d-t}). These already represent the probably of observing an afterpulse with any charge in each time bin. Two complications arise from using the TTNP histograms: shadowing and dark noise. \emph{Shadowing} describes the fact that the probability of observing a pulse at a time $t$
depends on the probability to have already observed a pulse before time $t$, since we disregard any pulses that arrive after the first follower pulse. Dark noise contributes unwanted pulses to the histograms and thus represents background in this measurement.

\begin{figure}[htbp] 
	\centering
     \includegraphics[width=0.8\textwidth]{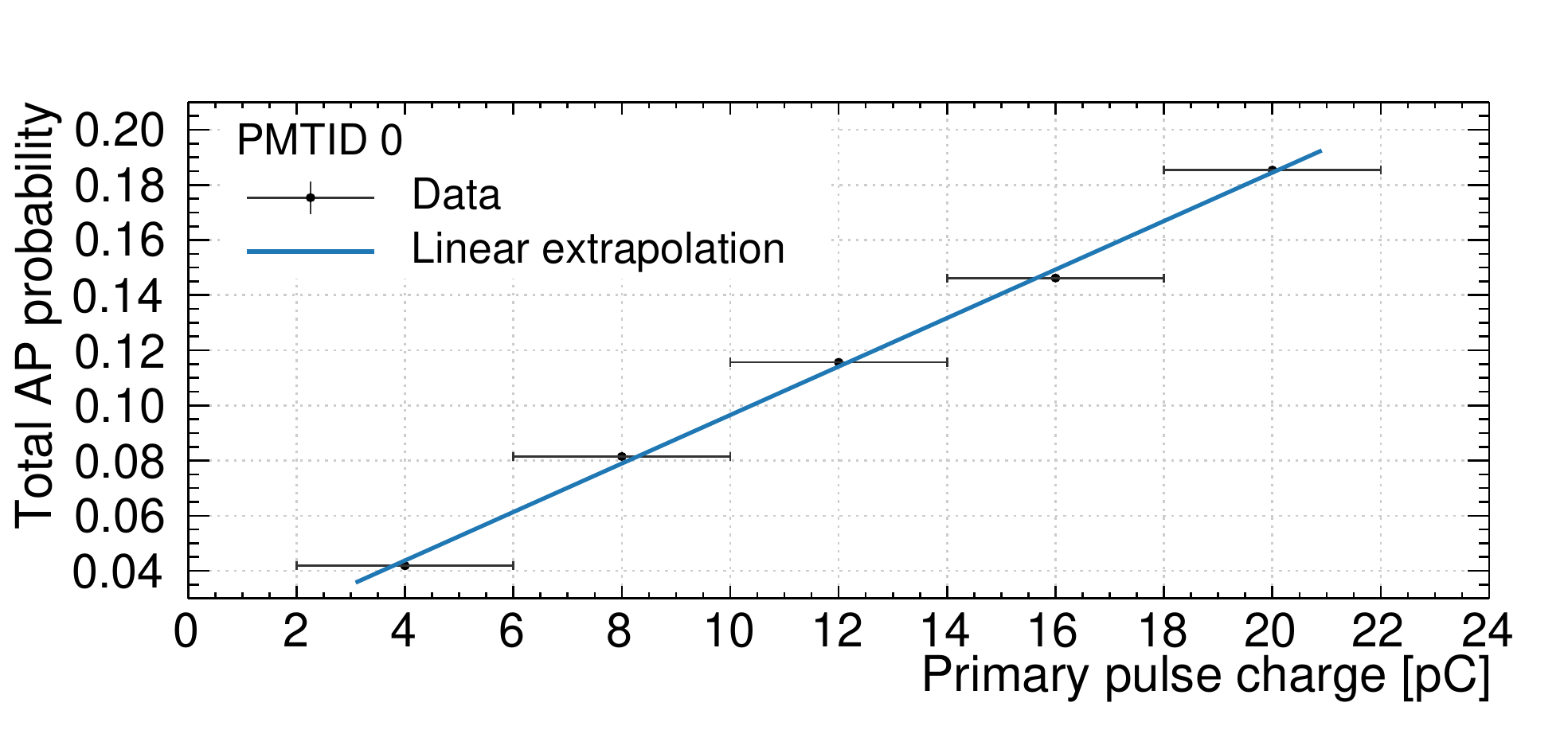}
     \caption[]{
       The total afterpulsing probability (given as a fraction, not as a percentage) as a function of the
       primary pulse charge is shown together with a linear fit. The grey horizontal bars indicate the charge intervals into which the primary pulses were sorted. The mean SPE charge for this PMT is 9.5~pC, and the fraction of primary pulses with more than 1~PE is smaller than 6\%. Statistical error bars are smaller than the marker size.}
     \label{fig:ap-1d-q}
\end{figure}

In order to calculate the total afterpulsing probability, the TTNP
distribution is first {\em unshadowed} by recalculating the
probability of observing an afterpulse at time $t$, and the uncertainty, according to the methodology described in \cite{retiere:2017}. In the limit that the
afterpulsing and dark noise probability in the time window in question
are small, the exact method cited approximates to
\begin{equation}
P(t) \simeq P^{\mbox{\tiny shadow}}(t) / [1-\int_0^tP(t)]
\end{equation}
where $P(t)$ is the unshadowed afterpulsing probability distribution and $P^{\mbox{\tiny shadow}}(t)$ the shadowed one (i.e. the distributions shown in Fig.~\ref{fig:ap-1d-t}).
The unshadowed time spectrum is corrected for the dark noise rate and
then integrated from \SI{0.1}{$\mu$s} to \SI{10}{$\mu$s}. The dark noise rate is estimated by averaging the unshadowed pulse
probability between \SI{30}{$\mu$s} and \SI{170}{$\mu$s}.  The dark noise rate estimated this way is a slight over-estimate as discussed in Sect.~\ref{sect:dn}. Taking the largest plausible uncertainty in the dark noise rate to be 30\% (based on the data in Fig.~\ref{fig:DNvsT} at \SI{280}{K}), we get a conservative systematic error on the afterpulsing probability of $3\cdot10^{-3}$, which can be compared to the total AP probability of approximately 0.1 at the peak of the SPE charge distribution as shown in the example in Fig.~\ref{fig:ap-1d-q}.

The approximate method was used to give intuition and
double check the formally correct calculations used herein. 
The numerical difference between the results from the two methods is negligible when considering afterpulse times up to \SI{10}{$\mu$s} after the primary pulse.

We observe that the afterpulsing probability is a
function of the charge of the primary pulse, with each afterpulsing
population increasing in rate with higher primary pulse charge. The
total afterpulsing probability for primary pulse charges in the ranges
of 2-6~pC, 6-10~pC, 10-14~pC, 14-18~pC and 18-22~pC is shown in
Fig.~\ref{fig:ap-1d-q} for a typical PMT.  The statistical
uncertainty dominates the uncertainty on the total afterpulsing
probability. The mean SPE charge for
this PMT is 9.5~pC; charges lower or higher than this occur due to the
width of the SPE charge distribution (see
e.g. Fig.~\ref{fig:highoccspe}). The fraction of primary pulses with
more than 1~PE is smaller than 6\%.  The relationship between primary pulse charge and afterpulsing probability was modelled with a linear fit to yield approximately
\begin{equation}
P_{AP} = 0.0083(11) + 0.00890(12) \cdot Q_{primary}[\mbox{pC}] \label{eq:approb}
\end{equation}
with $\chi^2/$NDF = 22/3.

As evident from the large reduced $\chi^2$, the linear model does not
describe the afterpulsing probability perfectly. We estimated the
uncertainty on afterpulsing probabilities obtained with the linear
model by carrying out the analysis independently for nine afterpulsing
runs taken over the course of \SI{7}{months}. The variation for each PMT from run-to-run was calculated by
comparing the afterpulse probability from the linear fit at 10~pC. The
average RMS for a single PMT over the 9 runs was 4\%, which we take as
an estimate of the uncertainty.

We verified that the linear model of afterpulsing probability also applies for each of the three distinct afterpulsing populations by themselves.

\begin{figure}[htbp] 
\centering
     \includegraphics[width=0.8\textwidth]{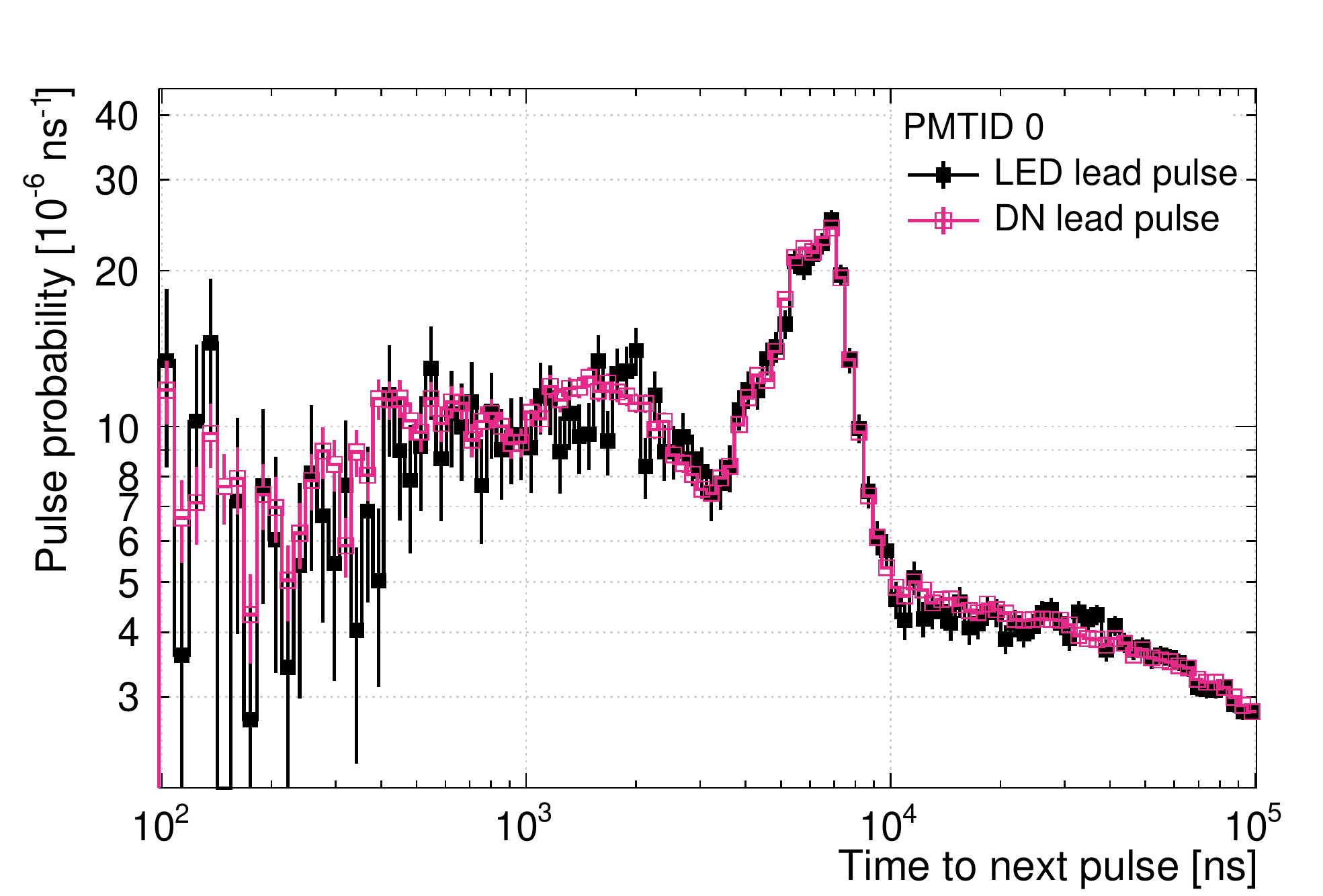}
     \caption[]{
       The time-to-next-pulse histogram is shown for primary pulses from injected light (LED) and from dark noise (DN), both with a charge range from 2 to 13~pC. The two datasets were taken on subsequent days and with all PMTs at room temperature.}
     \label{fig:apdarklight}
\end{figure}

The afterpulsing analysis described here was also carried out on a dataset where the primary pulse was
from dark noise rather than from injected light. Fig.~\ref{fig:apdarklight} shows the TTNP distributions for both types of primary pulse. The distributions are the same within uncertainty.

The AP distributions have similar features across the 255 DEAP PMTs. The average number and RMS (in brackets as percent of the average), across the PMT array, of PE per AP in each region is 0.6 (10\%), 2.0 (9\%), and 1.0 (3\%). Since the AP probabilities depend on the SPE charge, they drift if the the SPE charges drift. At the time of this calibration, the average AP probabilities in the three regions were 0.17\% (23\%), 2.3\% (22\%), and 5.7\% (26\%).

\subsection{Discussion}

Afterpulsing populations were found in three broad time regions
centered at 600~ns, 1500~ns and 6000~ns. The first population is
dominated by pulses with sub-SPE charges. One model for this behavior
is the positive ion striking the first dynode instead of the
photocathode and thus the anode signal would have one fewer amplification stage. The second population is dominated by multi-PE
afterpulses, while the third one has mainly SPE pulses. The charge and
time of an afterpulse are correlated in non-trival ways.

We found that the total afterpulsing probability, as well as each of the afterpulsing probabilities in the three individual time regions, have a near-linear correlation to the charge of the primary pulse, even though 94\% of the primary pulses are single PE.
This suggests that
the absolute charge, rather than the number of photoelectrons released from
the cathode, is the determining factor in the afterpulsing probability.

The run-to-run uncertainties have a relative variation of 4\%. The systematic uncertainty from subtraction of dark noise contributes a  3\% uncertainty. We assign therefore a 5\% uncertainty in afterpulsing rate.

Where not otherwise noted, the data presented in this section were taken
when the detector was partly cooled (to reduce dark noise from thermal
emission from the photocathode) but when the amount of argon in the
detector was small.

The afterpulsing probabilities were measured monthly for all PMTs over
a year of room-temperature detector operation and remained stable over this time
period. The afterpulsing probabilities also remained unchanged while the
PMTs cooled to \SI{\ensuremath{\simeq 5}}{C}.

\section{Measuring dark noise rates}\label{sect:dn}
Dark noise consists of individual SPE pulses observed at random times that can neither be attributed to photons from argon scintillation light nor to afterpulsing. We also count stray photons from e.g. Cherenkov light in the light guides as dark noise.
At room temperature and under normal bias voltage, photoelectrons released from the photocathode due to thermal fluctuations (thermionic emission) dominate the dark noise rate \cite{Hamamatsu:2007tc}. 

The dark noise rate for each PMT can be determined from three different methods. Methods one and two make use of the regular LED monitoring datasets. The LED flashes \SI{6.5}{$\mu$s} into the start of the waveform. The number of pulses within the first \SI{5}{$\mu$s} of each event ($N_p$), as well as the number of events with no pulses ($N_0$) in that time window are counted. Method one, based on the pulses observed, calculates the dark noise rate $R_{DN}$ as
\begin{equation}
R_{DN} = \frac{N_p}{N_{t} \; \Delta t}
\end{equation}
where $N_t$ is the total number of events and $\Delta t$ is the time window in which pulses are counted. This \emph{pulse-counting method} over-estimates $R_{DN}$ by an amount equal to the afterpulsing probability, because some of the pulses counted are afterpulses.

Method two is based on Poisson statistics. Given $R_{DN}$, the probability of observing no pulses in time window $\Delta t$ is
\begin{equation}
P_0 = e^{-R_{DN}\cdot \Delta t} = \frac{N_0}{N_t}
\end{equation}
thus
\begin{equation}
R_{DN} = - \frac{1}{\Delta t} \text{ln}\big(\frac{N_0}{N_t}\big) 
\end{equation}
This \emph{zero-pulse method} method also overestimates the dark noise rate, 
because an earlier dark noise pulse can produce a non-zero pulse event by producing an afterpulse in the measurement time window. The bias is smaller than in method one, because this method is insensitive to afterpulses that occur in the same time window as a dark noise pulse.

In method three, the pulse rate at times $t >10$~$\mu$s in the unshadowed TTNP distribution from the afterpulsing study (Sect.~\ref{sect:ap}) is taken as the dark noise rate. This method is unbiased as the analysis guarantees no afterpulses in this time-region, it is however not suited to ongoing monitoring of the PMTs because taking this type of data frequently is impractical.

\begin{figure}[htbp] 
\centering
     \includegraphics[width=0.8\textwidth]{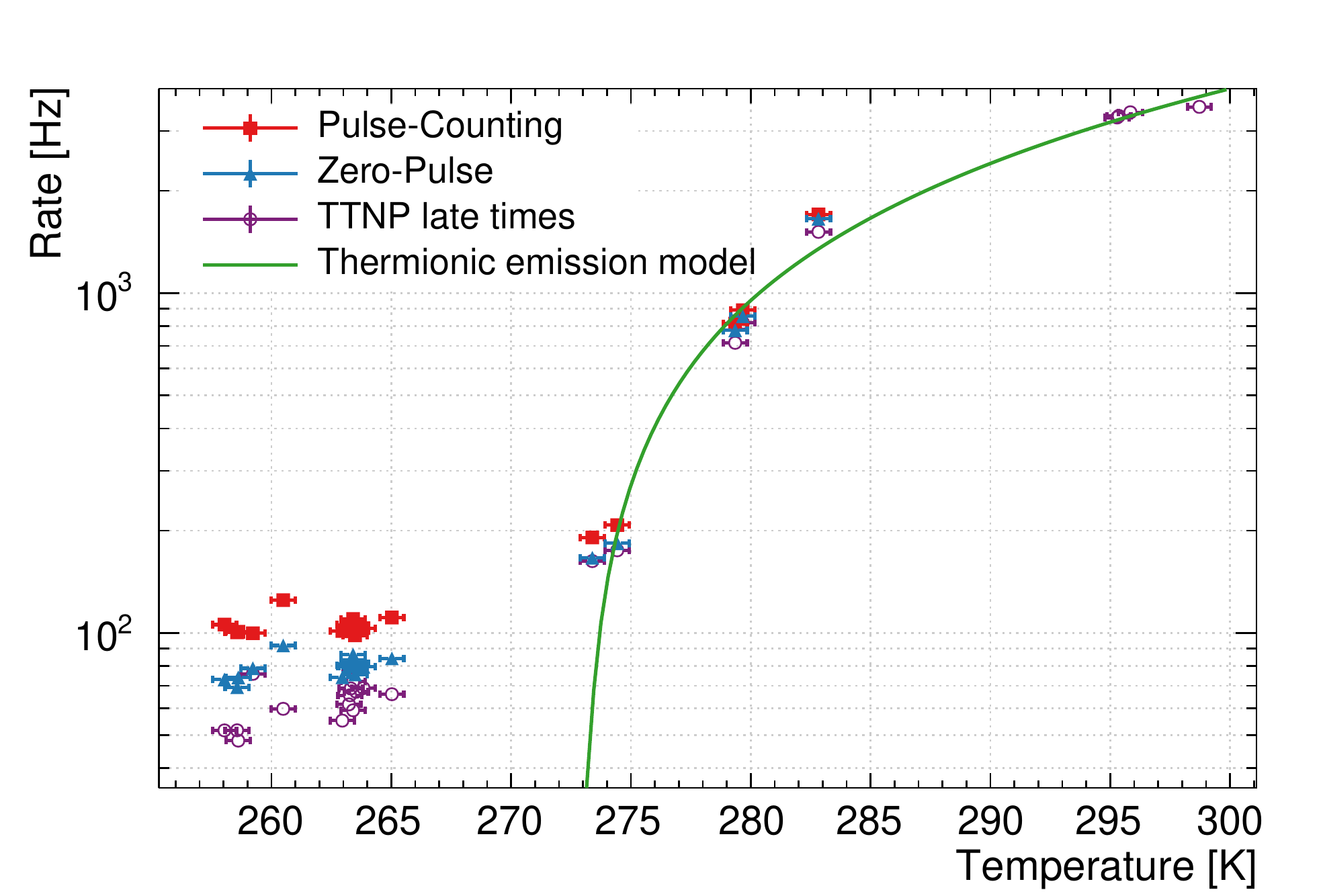}
     \caption[]{The dark noise rate for PMTID~253 versus PMT temperature, measured using the three different methods described in the text. The fit uses Eq.~\eqref{eq:dnbyt}. Error bars are statistical only, and the systematic shift between the analysis methods is due to some afterpulses being counted as dark noise in the pulse-counting and the zero-pulse methods. The PMTs operating temperature is approximately \SI{260}{K}.}
     \label{fig:DNvsT}
\end{figure}

Fig.~\ref{fig:DNvsT} shows the dark noise rate determined using each of the three methods for a representative PMT versus the PMT's temperature. The PMT temperature is measured by an RTD temperature probe attached to the copper thermal short near the PMT, and has a systematic uncertainty of about \SI{2}{K} from the RTD calibration. The actual temperature of the photocathode may be slightly different.

Following \cite{Meyer:2008qb}, we use Richardson's law of thermionic emission to describe the temperature dependence:
\begin{equation}
R_{DN}(T) = T^{2}\cdot e^{-W/kT}+\beta \label{eq:dnbyt}
\end{equation}
where $k$ is Boltzman's constant, $T$ is the PMT temperature, $W$ is the work function at the PMTs bias voltage, and $\beta$ is the saturation value of the dark noise rate at very high temperatures.

This model describes the data well above \SI{273}{K}. Going to lower temperatures, the rate falls slower than expected from thermionic emission. This behaviour at lower temperatures has been observed for various types of PMTs \cite{Meyer:2008qb} and can be attributed to other processes dominating the dark rate, such as scintillation and Cherenkov light from radioactivity in the PMT glass or surrounding material \cite{Hamamatsu:2007tc}.

\section{Conclusion}
We have shown the Hamamatsu R5912-HQE can be calibrated effectively for charge response, including the SPE charge distribution, double and late pulsing, afterpulsing, and dark rates. 


By developing a full model of the low-light charge distribution, we were able to fit the whole distribution instead of just the region near the SPE peak, as is commonly done. Doing so, we found that the SPE charge distribution requires an exponential component at low charges. This component has been observed in similar PMT types before and  shifts the true mean of the SPE distribution lower (9\% lower for the sample PMT discussed here) than one would obtain from considering only the region near the SPE peak, and thus affects PE counts. By studying charge response as a function of light intensity, threshold effects were studied and a systematic uncertainty on the SPE charge was found to be 3\%. 

Afterpulsing, late, and double pulsing affect the liquid argon pulse shapes the DEAP-3600 detector measures and thus influence PE counts and pulse shape discrimination methods. The measured afterpulsing time and charge distribution will not only allow accurate simulation of the observed signals, but also inform mitigation strategies such as excluding the dominant afterpulsing region from analysis, or statistically separating LAr scintillation pulses from afterpulses based on the respective probabilities to find a pulse of a certain size at the pulse time in the event. This is the first use of the technique in~\cite{retiere:2017} to measure afterpulsing of PMTs. 

Doing afterpulsing, late, and double pulsing measurements as described here was easier during detector commissioning; once the DEAP-3600 detector filled with LAr, the $^{39}$Ar coincidence rate of over \SI{3}{kHz} makes measurements more challenging. The possibility to obtain afterpulsing rates from the LAr pulse shapes is under investigation. 

The dark noise rates have the expected strong temperature dependence. A measurement of the true dark noise rate is challenging when the detector is filled, again due to the large $^{39}$Ar coincidence rate. However, most analyses only require an effective dark rate, which is the overall probability of observing a pulse in an event that did not come from the event itself, be it due to dark noise, $^{39}$Ar scintillation photons, or other sources. This rate is naturally obtained from either of the two dedicated dark noise analysis methods presented here.

The PMT behaviour under multi-PE signals, such as the linearity of the afterpulsing probability with incident PE, and the linearity of the charge response, as well as the SPE pulse shape and PMT efficiencies, will be studied in future work.

\section*{Acknowledgements}
This work was funded by the Natural Sciences and Engineering Research Council of Canada, the Canadian Foundation for Innovation, the Ontario Ministry of Research and Innovation, the Alberta Advanced Education and Technology Ministry (Alberta Science and Research Investments Program), the European Research Council Project ERC StG 279980, the UK Science \& Technology Facilities Council (STFC) grant ST/K002570/1, the Leverhulme Trust grant number ECF-20130496, the Rutherford Appleton Laboratory Particle Physics Division, STFC and SEPNet PhD studentship, and DGAPA-UNAM through grant PAPIIT No. IA100316.
We thank SNOLAB and its staff for support through underground space, logistical and technical services. SNOLAB operations are supported by the Canada Foundation for Innovation and the Province of Ontario Ministry of Research and Innovation, with underground access provided by Vale at the Creighton mine site.
We thank Compute Canada, Calcul Qu\'ebec, and the Center for Advanced Computing at Queen's University (HPCVL) for the excellent computing resources and support they provided to undertake this work. The work of many summer and co-op students is gratefully acknowledged. 

\bibliography{pmtpaper}

\end{document}